%% file: Bcfrac-prl-v5.tex
\pdfoutput=1

\documentclass[12pt,a4paper]{article}
\usepackage[bottom,flushmargin,hang,multiple]{footmisc}
\usepackage{ifthen} 
\newboolean{pdflatex}
\setboolean{pdflatex}{true} 

\newboolean{articletitles}
\setboolean{articletitles}{true} 

\newboolean{uprightparticles}
\setboolean{uprightparticles}{false} 

\def\paperauthors{LHCb collaboration} 
\def\paperasciititle{Measurement of the Bc meson production fraction and asymmetry in 7 and 13 TeV pp collisions} 
\def\papertitle{Measurement of the $B_c^-$ meson production fraction and asymmetry in 7 and 13\tev $pp$ collisions} 
\def\paperkeywords{{High Energy Physics}, {LHCb}} 
\def\papercopyright{\the\year\ CERN for the benefit of the LHCb collaboration} 
\def\paperlicence{CC-BY-4.0 licence}
\def\paperlicenceurl{https://creativecommons.org/licenses/by/4.0/}

\def\Bcmu  {\ensuremath{\Bcm\to\jpsi\mun\overline{\nu}}\xspace}
\def\Bcpsi  {\ensuremath{\Bcm\to\psitwos\mun\overline{\nu}}\xspace}
\def\Bctau  {\ensuremath{\Bcm\to\jpsi\taum\neut}\xspace}

\def\Bcchi  {\ensuremath{\Bcm\to\chi_{c0,1,2}\mun\overline{\nu}}\xspace}

\def\mcor  {\ensuremath{m_{\mathrm{cor}}}\xspace}

\def\araw  {\ensuremath{a_{\mathrm{raw}}}\xspace}
\def\aprod  {\ensuremath{a_{\mathrm{prod}}}\xspace}

\input{preamble}
\usepackage{longtable} 

\begin{document}

\renewcommand{\thefootnote}{\fnsymbol{footnote}}
\setcounter{footnote}{1}

\input{title-LHCb-PAPER}


\renewcommand{\thefootnote}{\arabic{footnote}}
\setcounter{footnote}{0}



\pagestyle{plain} 
\setcounter{page}{1}
\pagenumbering{arabic}



\input{1introduction}

\input{2bfprediction}

\input{detector}

\section{Event selection, signal efficiencies and yields}
\label{sec:Event}
\subsection{\boldmath Selection of $\Bcm\to\jpsi\mu^-\overline{\nu}$ candidates}
The analysis is done separately for the light $B$ meson modes and the \Bcmu decay. In each case the triggered events are subject to further filtering requirements. In addition, the $\jpsi\mu^-$ sample is subjected to a boosted decision tree (BDT), a multivariate classification
method, using the TMVA toolkit \cite{Hocker:2007ht,*TMVA4}.  This is not necessary for the $D^0$ or $D^+$ modes because they have large signals and are relatively free from backgrounds \cite{LHCb-PAPER-2018-050}.

For the $\jpsi\mu^-\overline{\nu}$ final state the initial selection requires that muons that satisfy the \jpsi candidate trigger each have minimum $\pt  > 550$~\!MeV, have large impact parameters with the PV, form a good quality vertex, have a reasonable flight distance significance from the PV, and have a summed $\pt > 2$~\!GeV. The ``companion" muon that is not part of the \jpsi decay must be well identified and form a good quality vertex with the \jpsi candidate, which must be downstream of the PV. 

To suppress muon tracks that are reconstructed more than once, we require a small minimum opening angle between the muons from the \jpsi decay and the companion muon momentum measured in the plane transverse to the beam line. Specifically, this opening angle must be greater than 0.8$^{\circ}$.
 The invariant mass
of the companion muon and the oppositely charged muon from $\jpsi$ must differ
from the known value of the $\jpsi$ mass by more than $50\mev$ \cite{PDG2018}, while the
invariant mass with the same charged muon is required to be larger than $400\mev$. 

\begin{figure}[htb]
  \begin{center}
    \includegraphics[width=0.45\linewidth]{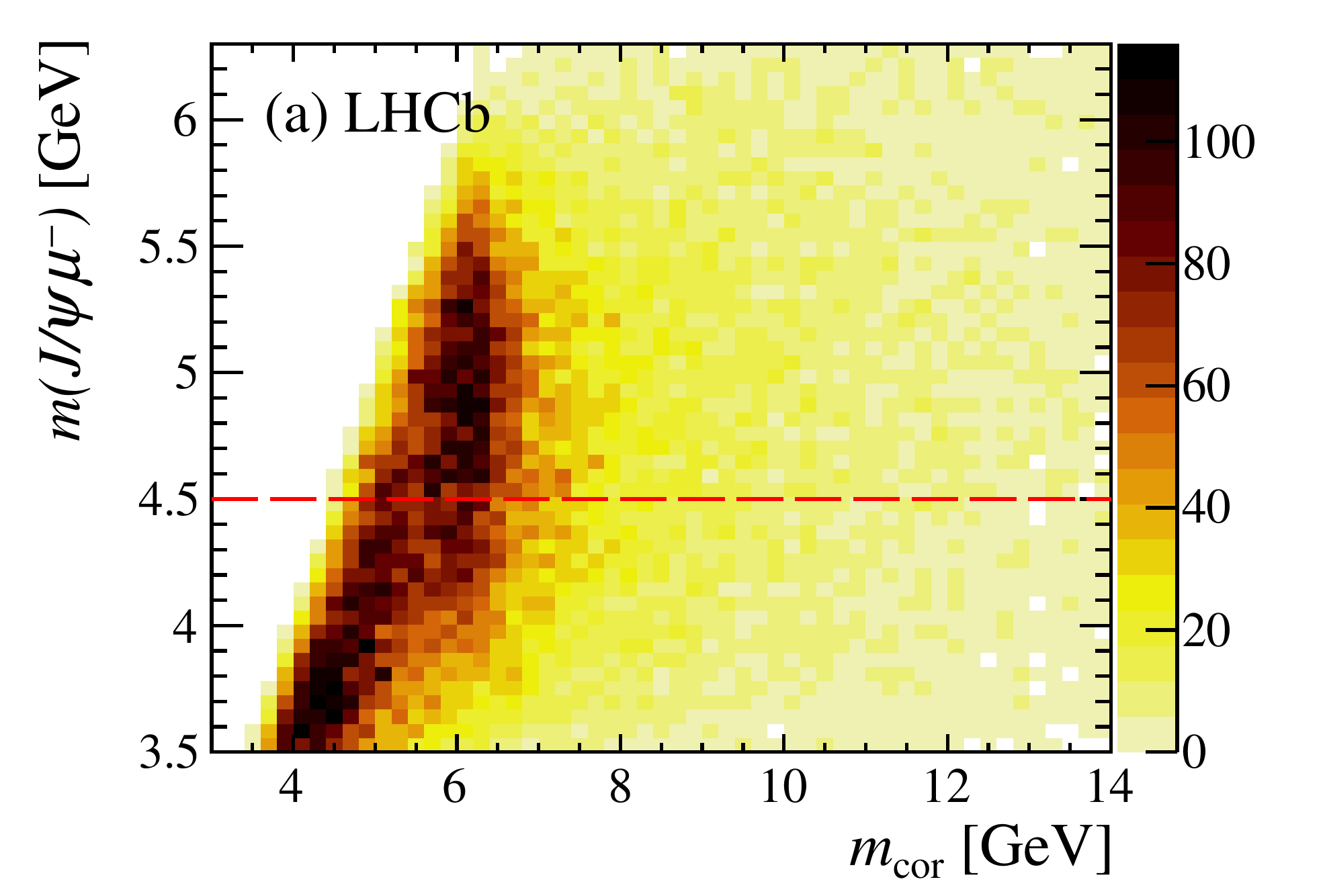}
    \includegraphics[width=0.45\linewidth]{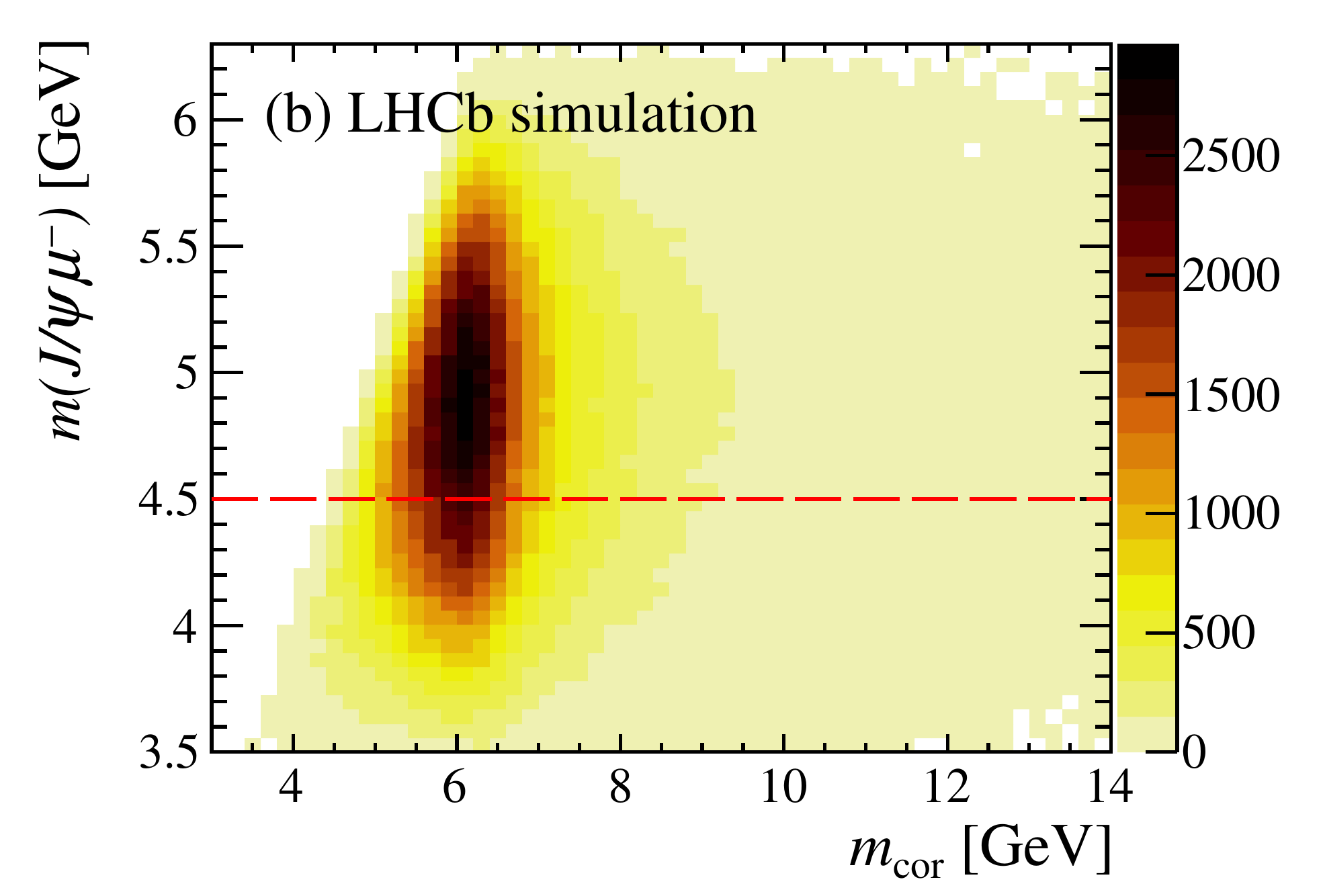}\\
    \includegraphics[width=0.45\linewidth]{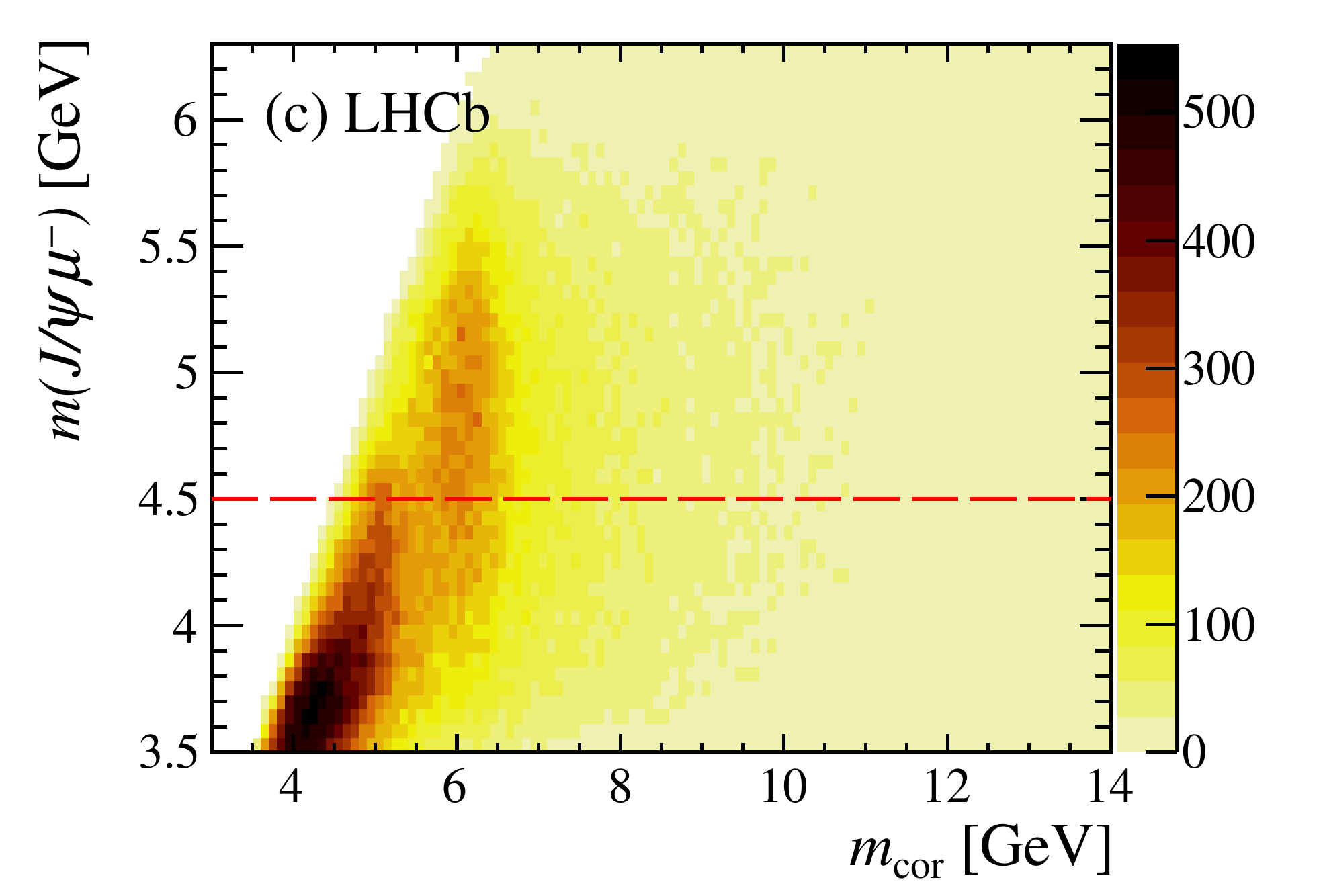}
    \includegraphics[width=0.45\linewidth]{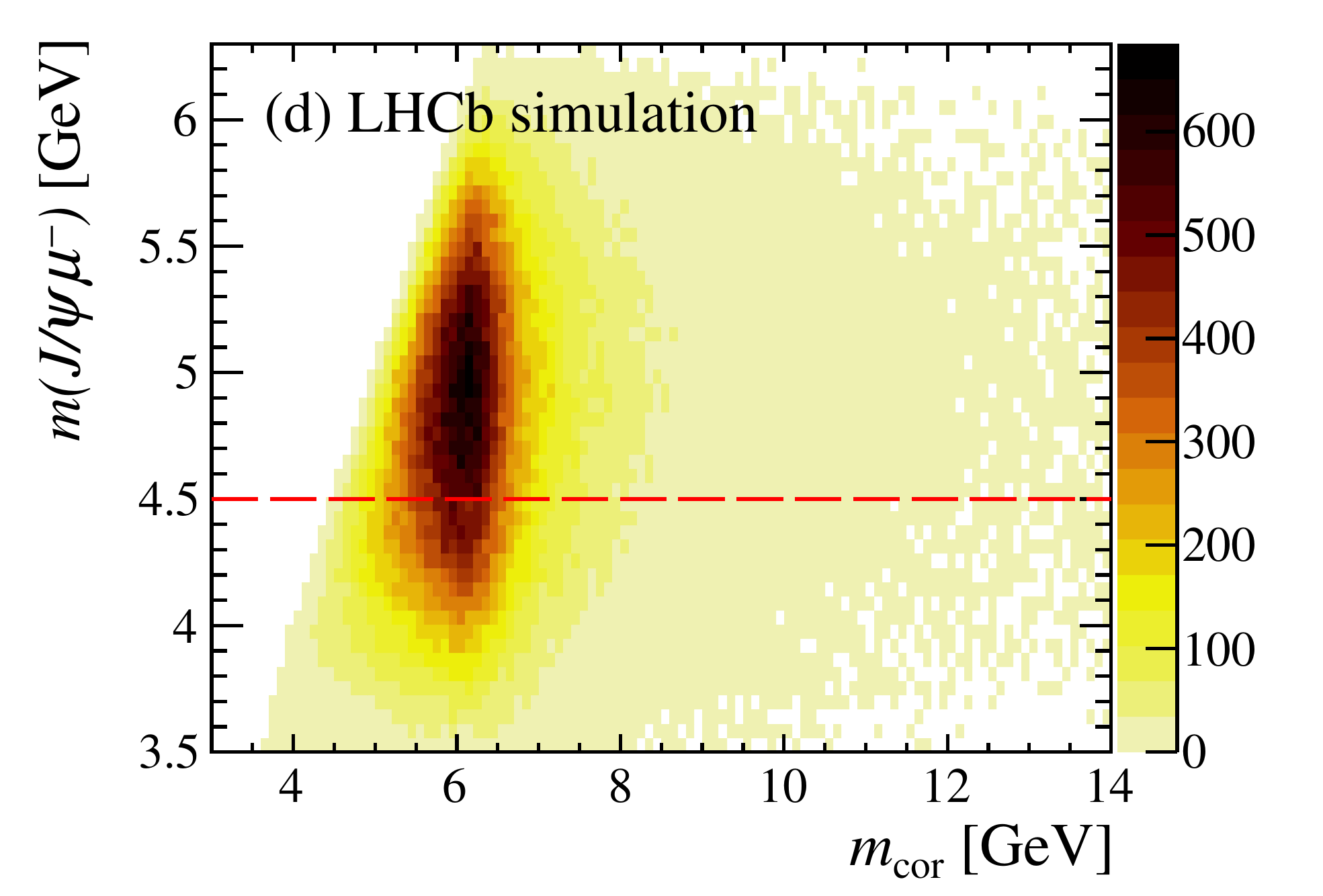}\\
    \vspace*{-0.5cm}
  \end{center}
  \caption{Distributions of corrected mass $\mcor$ and $m(\jpsi\mu^-)$ for (top) $7\tev$  and
    (bottom) $13\tev$ data, where (a) and (c) are data and (b) and (d) simulated signal. The (red) dashed line indicates the  $m(\jpsi\mu^-)>4.5\gev$ requirement. \label{fig:BcInvMassCut}
  }
\end{figure}

Since we are dealing with an exclusive final state, we define 
\begin{equation}
m_{\rm cor} \equiv \sqrt{m(\jpsi\mu^-)^{2} + p_{\perp}^{2}} + p_{\perp},
\end{equation}
where $p_{\perp}$ is the magnitude of the combination's momentum component transverse to the $b$-hadron flight direction.  Figure~\ref{fig:BcInvMassCut} shows the distributions of $m_{\rm cor}$ versus the invariant $\jpsi\mu^-$ mass, $m(\jpsi\mu^-)$, for both data and simulation. 
 To remove 
background, a requirement of $m(\jpsi\mu^-)>4.5\gev$ is applied, as indicated by the (red) dashed
line.

Since we also measure the production asymmetry between $B_c^+$ and $\Bcm$ mesons, we restrict the angular acceptance of the companion muon to make it more uniform by removing muons close to the edge of the detector, in the bending direction ($x$-direction), where large acceptance-induced asymmetries can occur. Thus, we require that the $x$-component of the momentum satisfies
\begin{equation}
   |p_x|\leq 0.294(p_z-2\gev),
\end{equation}
where $p_z$ is the muon momentum along the direction of the proton beam downstream of the PV,
as is done in Refs.~\cite{LHCb-PAPER-2016-051,LHCb-PAPER-2016-054}.

After these initial restrictions, we turn to the multivariate selection, forming the classifier denoted BDT in the following. The discriminating variables used  are: (a) the $\chi^2$ of the vertex fit of the \jpsi with the $\mu^-$; (b) the $\ln\chisqip$,
where \chisqip\ is defined as the $\chisq$ of the impact parameter with respect to the PV, of the $\jpsi$, $\mu^-$ and their combination; (c) the \pt of the \jpsi and the $\mu^-$;  and (d) the cosine of the angle between the $\mu^-$ and the \jpsi meson in the plane perpendicular to the beam direction. The training sample for signal is simulated $\Bcm\to\jpsi\mu^-\overline{\nu}$ events, and for background is inclusive $b\to \jpsi X$ simulated events.

We then optimize the BDT output threshold by maximizing $S/\sqrt{S+B}$, where $S$ and $B$ are the number of the signal and background yields in the signal region defined as
$\mcor\in(4.8,10.8)\gev$. The sum, $S+B$, is the total number of events within these limits, and $S$ is taken from a  fit to the $\mcor$ distribution. The optimal BDT output threshold  results in a BDT signal efficiency of  89\% with a background rejection of 63\%, as determined by observing the resulting samples of input signal simulation events and background candidates.
 
\begin{figure}[t]
  \begin{center}
    \includegraphics[width=0.45\linewidth]{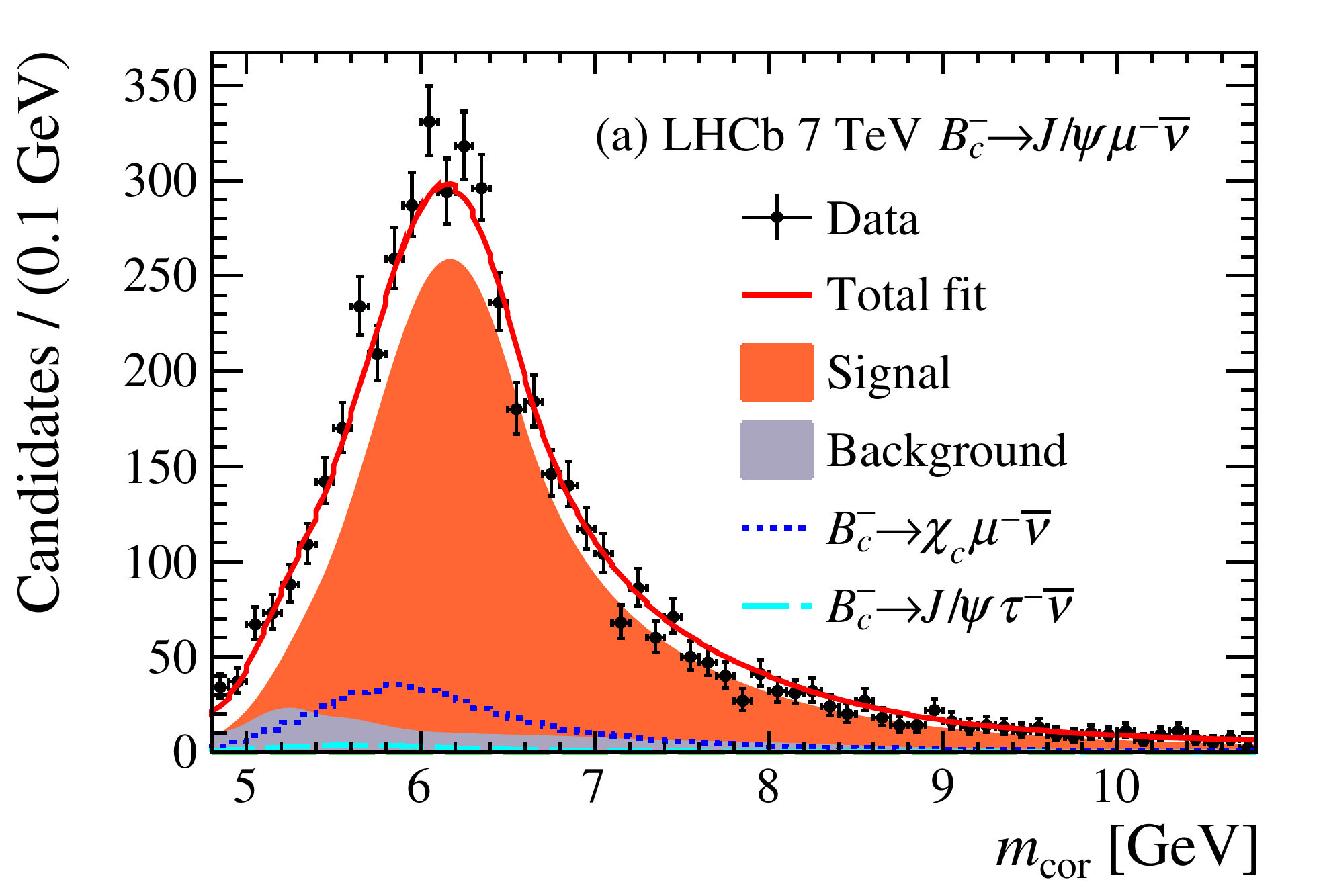}
    \includegraphics[width=0.45\linewidth]{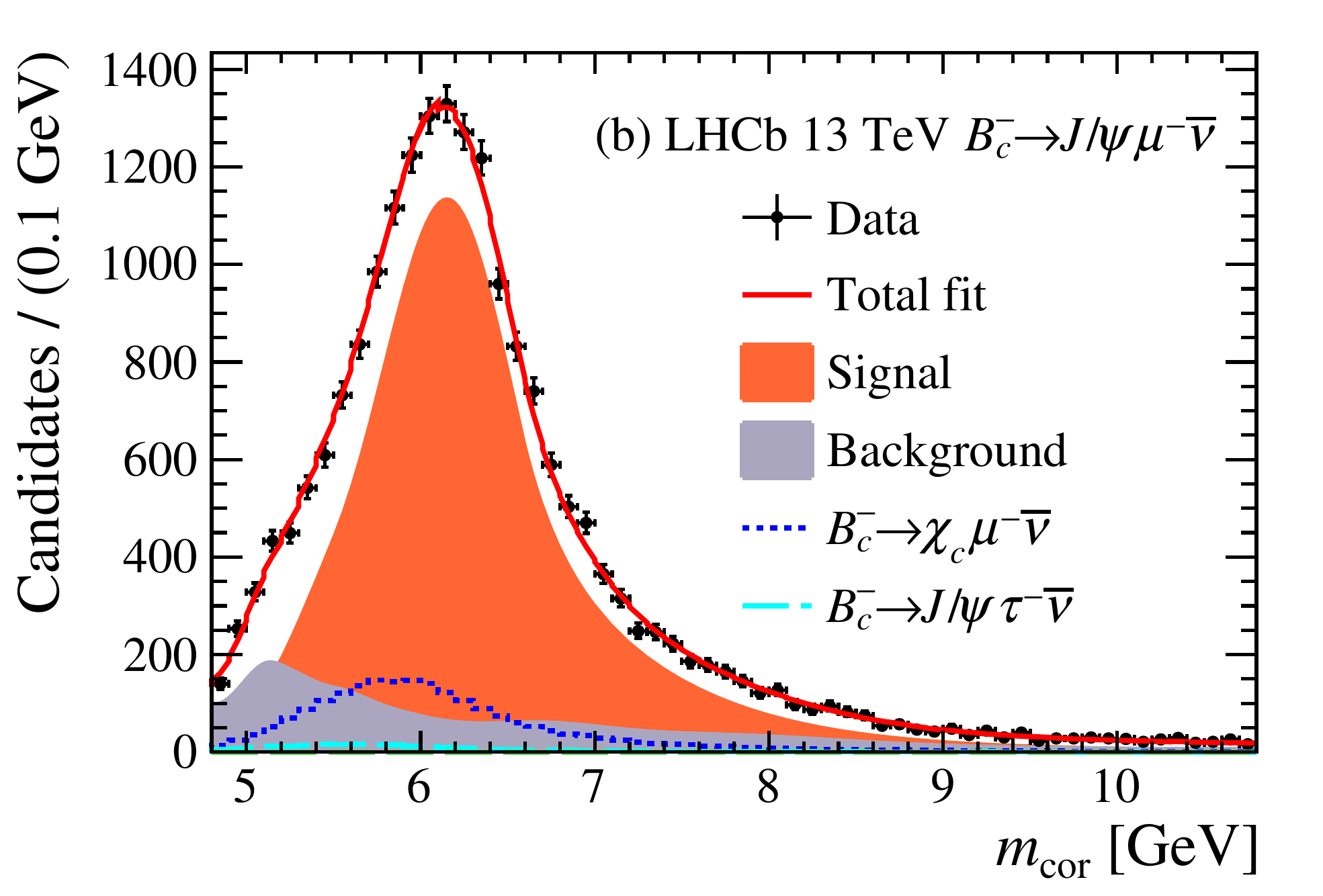}\\
    \vspace*{-0.5cm}
  \end{center}
  \caption{
     Fitted $\mcor$ distributions in (a) $7\tev$
      and (b) $13\tev$ samples. The signal and the backgrounds are shown as the dark (orange) and lighter (gray) areas.
 The dashed (cyan) curves show the $\Bctau$ components, while the dotted (blue) curves show the $\Bcchi$ components.
 The  $\Bcpsi$ contribution is also in the fit but is too small to be seen. The total fit is shown by the solid (red) curve. 
  }
  \label{fig:yield_data_all}
\end{figure}

The $\mcor$ distribution is shown in Fig.~\ref{fig:yield_data_all}. It consists not only of signal $\Bcm$ events, but also of $B_c^-\to \jpsi \tau^-\overline{\nu}$ decays, where $\tau^-\to\mu^-\nu\overline{\nu}$, and other $c\cquarkbar$ final states, most importantly  $\Bcpsi$ and $B_c^-\to \chi_c \mu^-\overline{\nu}$. We find shapes for these final states using simulation. 
The signal shape is a sum of a double Crystal Ball and a bifurcated Gaussian functions. The sum of the combinatorial and misidentification backgrounds are represented by a Gaussian kernel shape \cite{Cranmer:2000du}. For the other background modes, we use histograms directly.
These shapes are fitted to the $\mcor$ distributions in Fig.~\ref{fig:yield_data_all} in order to determine the $\Bcmu$ yields. The ratio of the $\jpsi\tau^-\overline{\nu}$ yield to the $\jpsi\mu^-\overline{\nu}$ yield is fixed, after accounting for the relative detection efficiencies, from the LHCb measurement of $0.71\pm0.17\pm0.18$, where the first uncertainty is statistical and the second systematic \cite{LHCb-PAPER-2017-035}; this convention is used throughout this paper.  The other components of the fit are allowed to vary.  We find $4010\pm200$ and $15~\!170\pm710$ signal $\Bcmu$ events at 7 and 13\tev, respectively, while the backgrounds sum to 950 and 5170 events at the same energies. These signal yields need to be corrected for the small background from candidates with a correctly reconstructed \jpsi meson that is paired with a hadron mis-identified as a muon.

\subsection{\boldmath Efficiency for $\Bcmu$}
\begin{figure}[b]
  \begin{center}
    \includegraphics[width=0.45\linewidth]{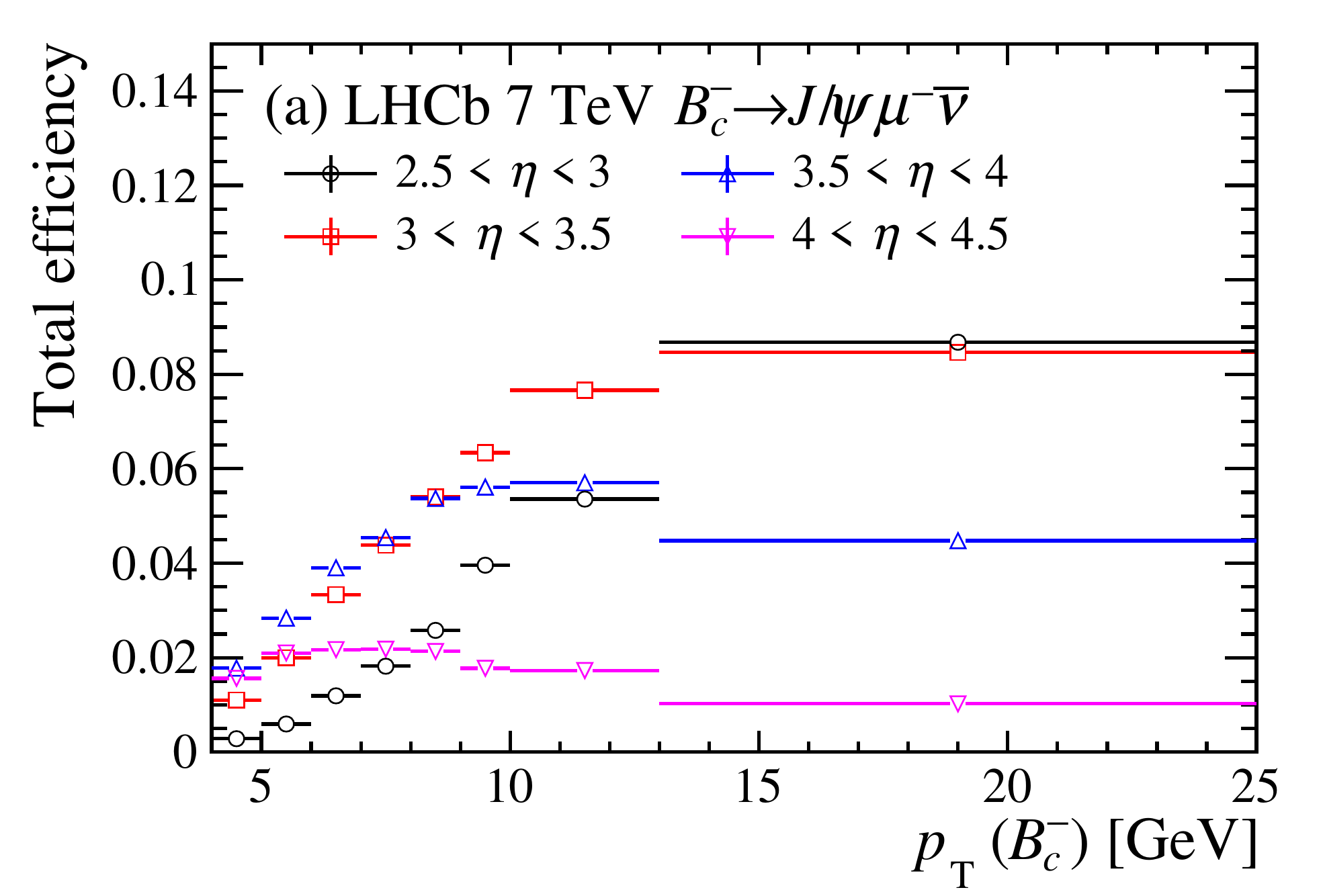}
    \includegraphics[width=0.45\linewidth]{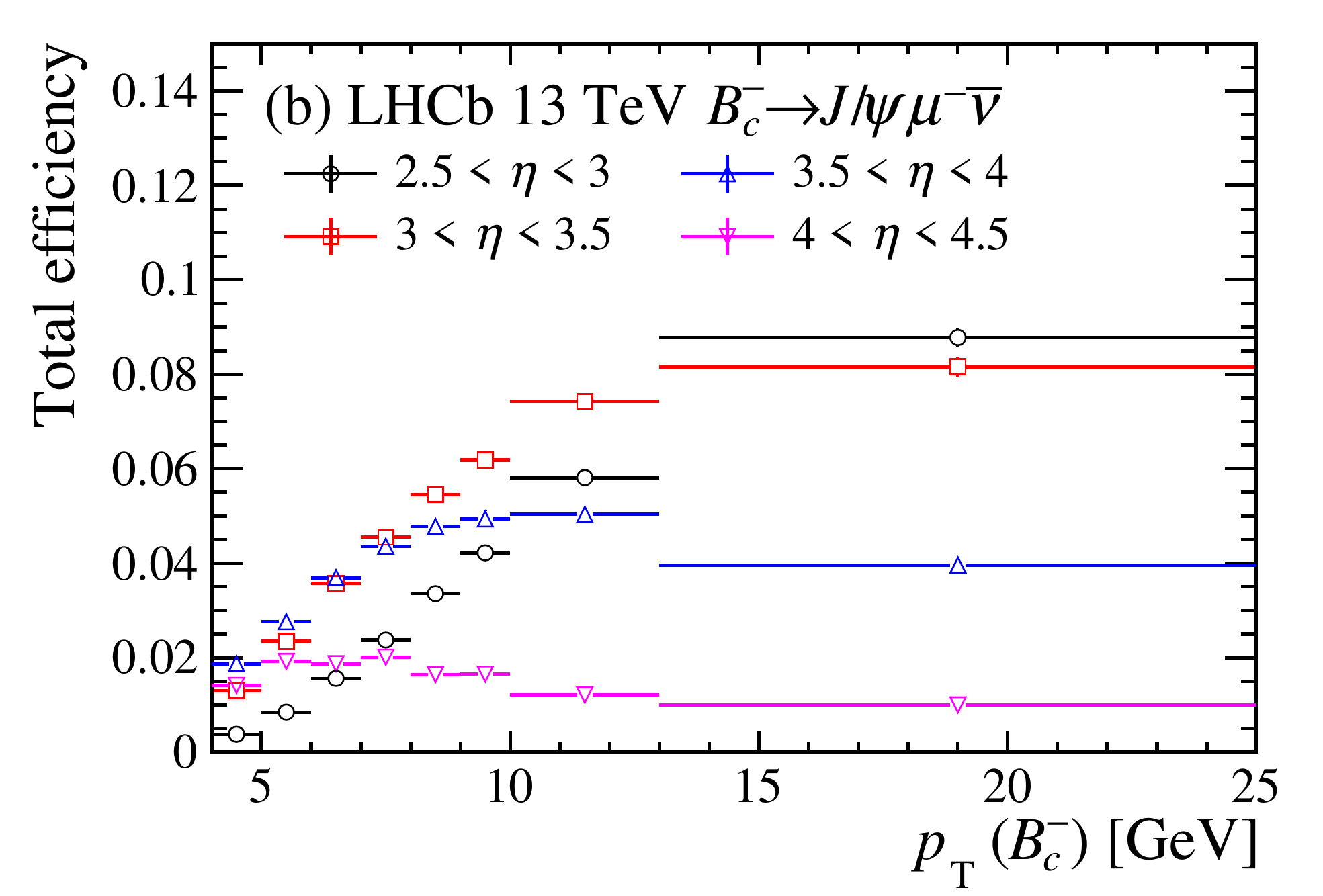}
    \vspace*{-0.5cm}
  \end{center}
  \caption{
     The total efficiency for  $\Bcmu$, as a function of $\pt(\Bcm)$ in different intervals of $\eta$ 
     in (a) 7 TeV and (b) 13 TeV samples.
  }
  \label{fig:efficiency_total}
\end{figure}

Efficiencies are determined using both data \cite{LHCb-DP-2018-001,LHCb-DP-2013-002} and simulation of $\Bcmu$, with the generated events weighted to match the $\pt(H_b)$, and $\eta$ distributions observed in data. In addition, we weight accordingly the $\chisqip$ distribution of the muon associated with the \jpsi. Weighting the simulation is important since
the total efficiencies are functions of these variables. Efficiencies using data include trigger, and muon identification. Efficiencies using simulation include detector acceptance, reconstruction and event selection, and removal of beam crossings with an excess number of hits in the detector.
Total efficiencies as a function of $\pt(\Bcm)$ for different $\eta$ intervals are shown in Fig.~\ref{fig:efficiency_total}.

\subsection{\boldmath $H_c X\mu^-\overline{\nu}$ selection criteria}
Selection criteria for $H_b\to H_c X \mu^-\overline{\nu}$ final states differ from those containing a $\jpsi$. The transverse momentum of each hadron must be greater than 0.3\gev, and that of the muon larger than 1.3\gev. 
We require $\chisqip >9$ with respect to any PV, ensuring that tracks do not originate from primary $pp$ interactions.
 All final state particles are required to be positively identified using information from the RICH detectors. Particles from $H_c$ decay candidates must have a good fit to a common vertex with $\chi^2$/ndof $<9$, where ndof is the number of degrees of freedom. They must also be well separated from the nearest PV, with the flight distance divided by its uncertainty greater than 5. 

Candidate $b$ hadrons are formed by combining $H_c$ and muon candidates originating from a common vertex with $\chi^2$/ndof $<9$ and an $H_c\mu^-$ invariant mass in the range 3.0--5.0\gev. 

Background from prompt $H_c$ production at the PV needs to be considered. We use the natural logarithm of the $H_c$ impact parameter, IP, with respect to the PV in units of mm.
 Requiring  ln(IP/mm)$>-3$ is found to reduce the prompt component to be below 0.1\%,  while preserving 97\% of all signals. This restriction allows us to perform fits only to the $H_c$ candidate mass spectra to find the $b$-hadron decay yields.

\begin{figure}[b]
  \begin{center}
    \includegraphics[width=0.45\linewidth]{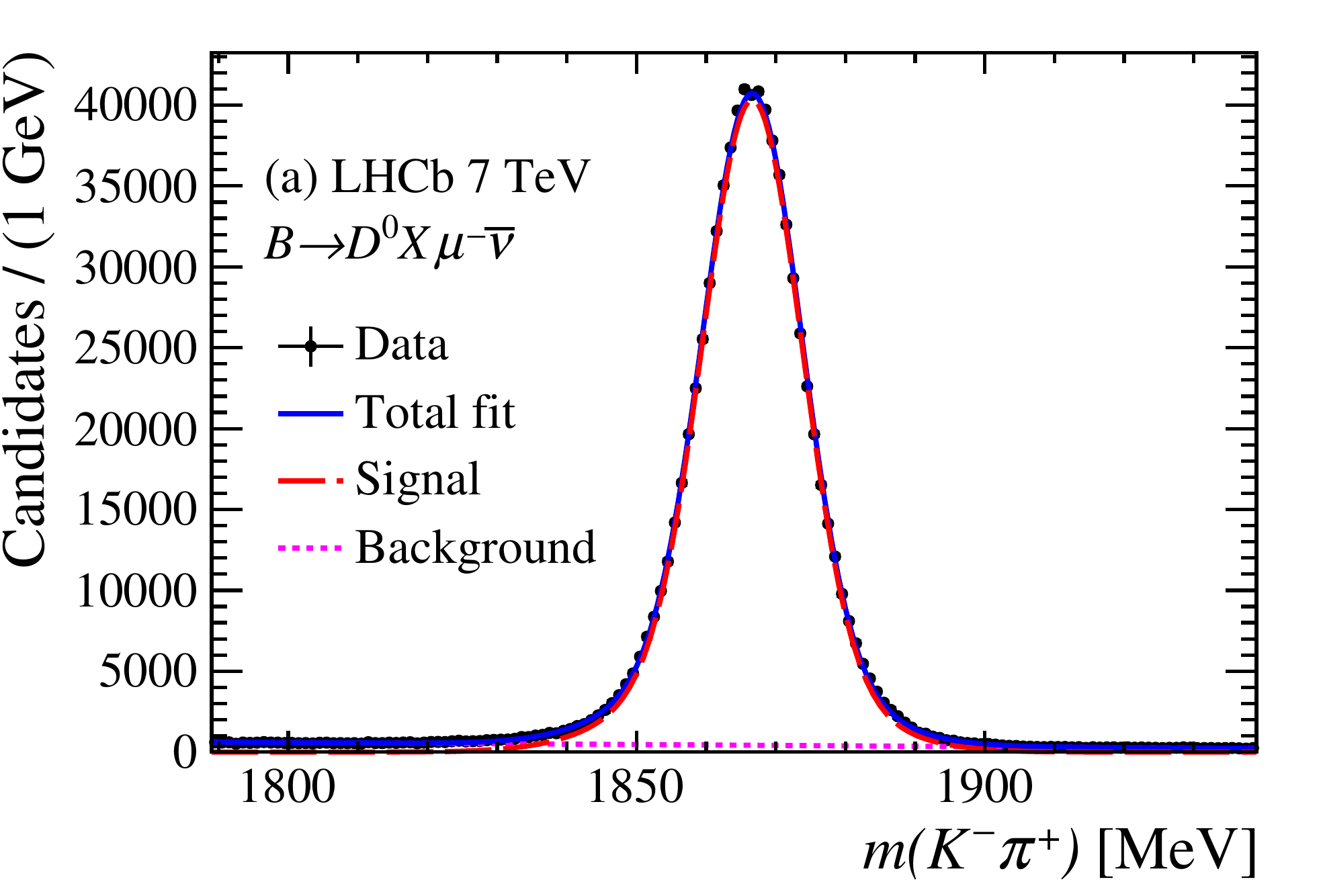}
    \includegraphics[width=0.45\linewidth]{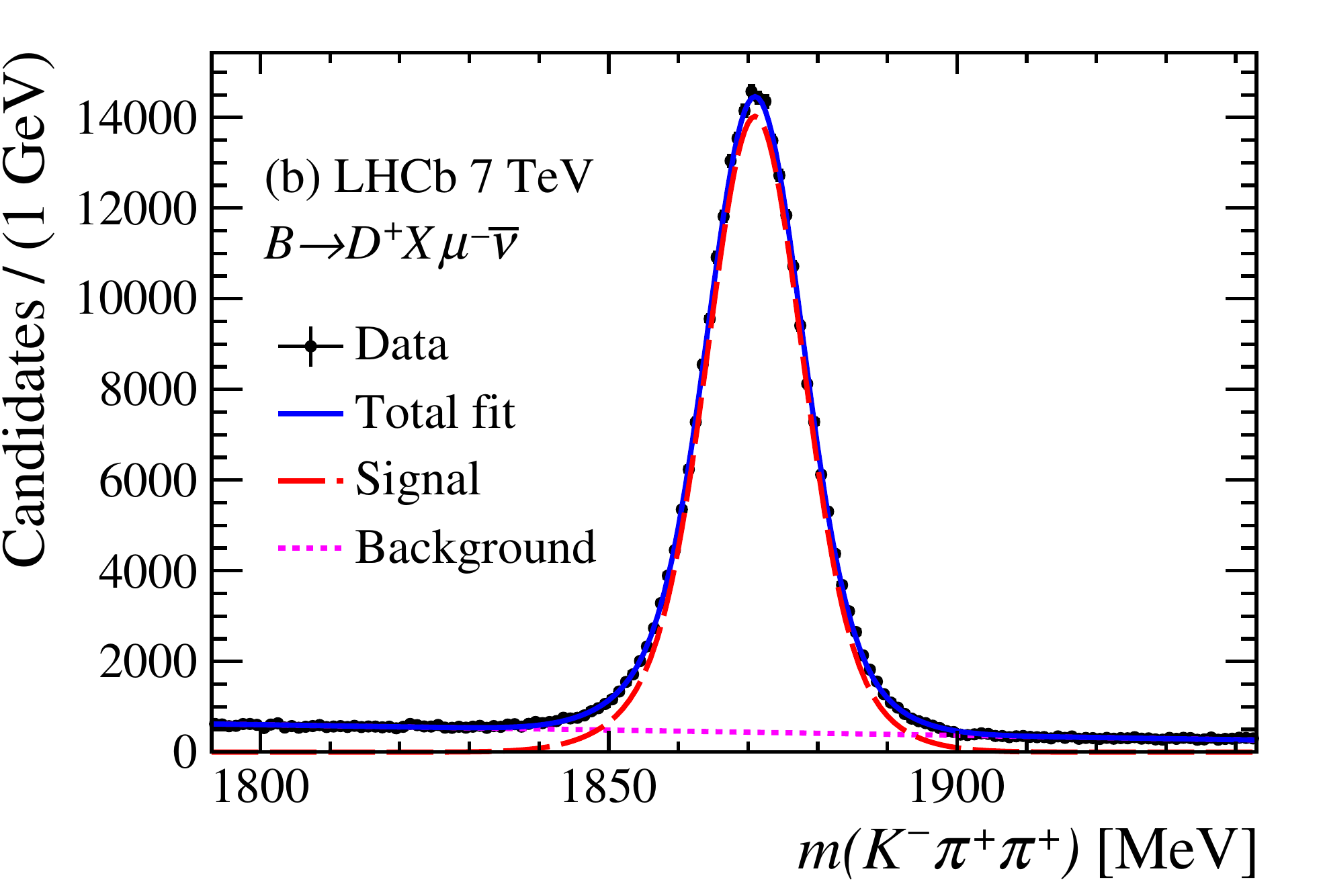}
    \includegraphics[width=0.45\linewidth]{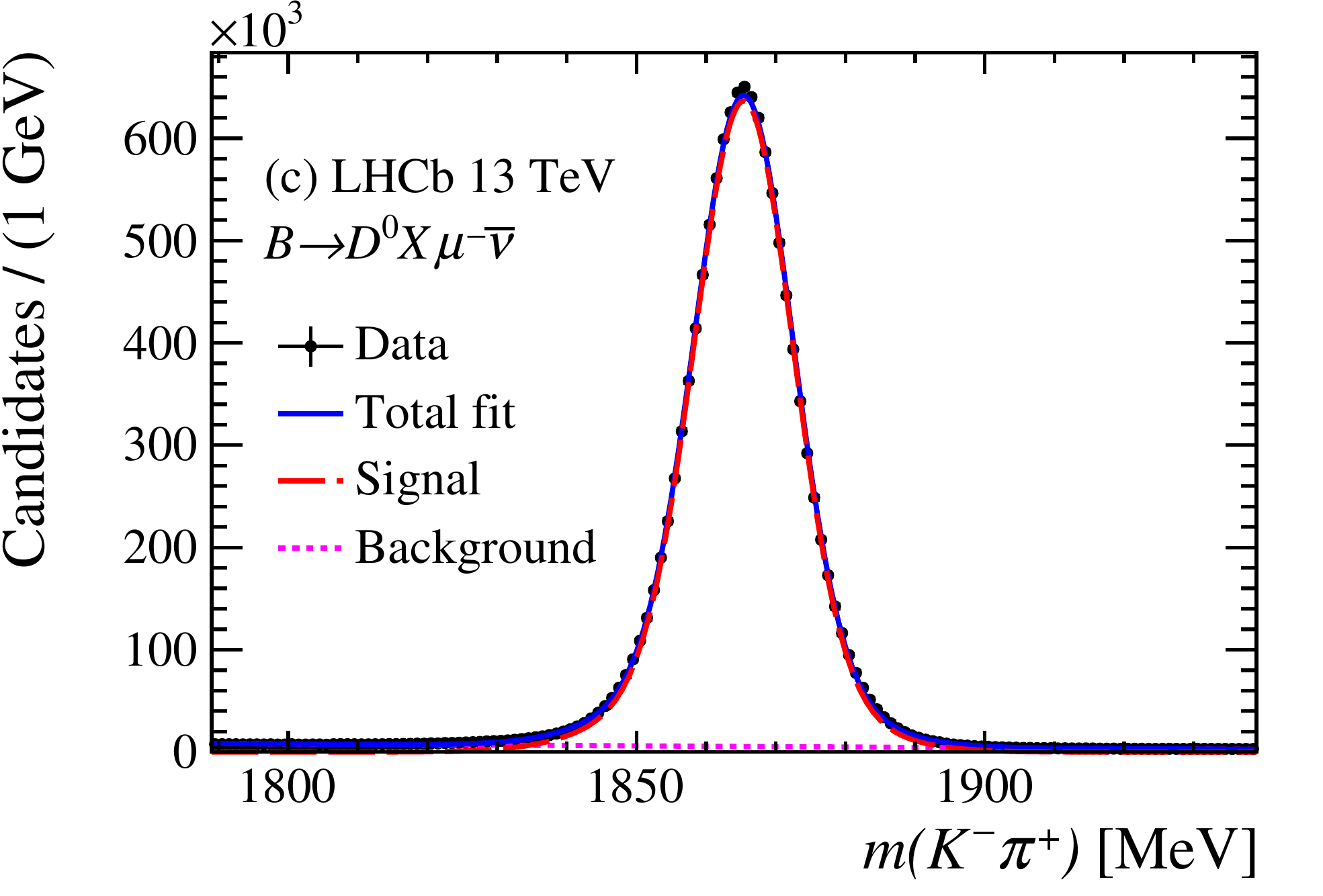}
    \includegraphics[width=0.45\linewidth]{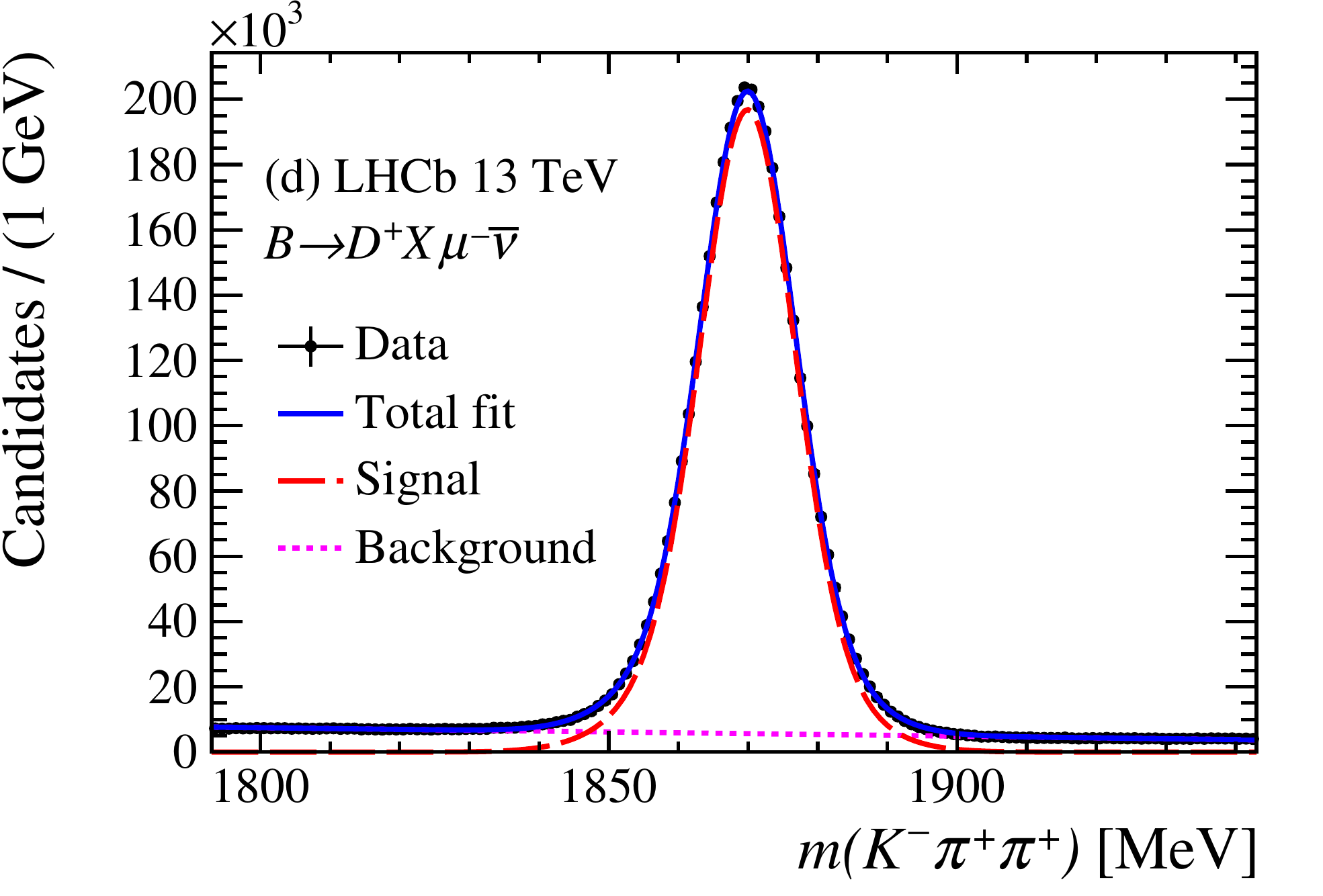}
    \vspace*{-0.5cm}
  \end{center}
  \caption{
     Invariant-mass distributions of (a) $K^-\pi^+$ and (b) $K^-\pi^+\pi^+$ for 7\tev, and (c) and (d) for 13\tev collisions. The data are shown by solid points. The (red) dashed lines represent the signal components. The combinatorial backgrounds are shown as the dotted (magenta) line, and the solid (blue) line shows the total fit.
  }
   \label{fig:SignalYieldsBsl}
\end{figure}

The $H_c$ candidate mass distributions integrated over $\pt(H_b)$ and $\eta$ are shown in  Fig.~\ref{fig:SignalYieldsBsl}  and consist of a prominent peak resulting from signal, and a small contribution due to combinatorial background from random combinations of particles that pass the selection. They are fit with a signal component comprised of two Gaussian functions, and a combinatorial background component modeled as a linear function.  The fitted yields are listed in Table \ref{tab:SignalYields_BslTotal}. These numbers must be corrected for hadrons that are mis-identified as muons, and for semileptonic decays of \Bsb and \Lb hadrons that produce $D^0$ and $D^+$ mesons.

\begin{table}[t]
\caption{Yields of $B\to DX\mu^-\overline{\nu}$ decays.}
\begin{center}   \begin{tabular}{ccccc}
      \hline\hline
      Mode &\multicolumn{2}{c}{7\tev Yields}  & \multicolumn{2}{c}{13\tev yields}\\\hline
                    &  Signal &  fake muons & Signal & fake muons \\
 $\Dz X \mu\overline{\nu}$ & $789~\!800 \pm940$ & $5500\pm 160$ & $12~\!285~\!000\pm 3700$ &$115~\!155\pm 580$  \\
   $\Dp X\mu\overline{\nu}$ & $263~\!190\pm 570$ & $~~\!990\pm ~70$& ~$3~\!686~\!240 \pm 2130$& ~$21~\!370\pm 240$\\
       \hline\hline
   \end{tabular}\end{center}
   \label{tab:SignalYields_BslTotal}
\end{table}

In Table~\ref{tab:SignalYields_BslTotal} the column labeled ``fake muons" shows the yields of wrong-sign $D^0X\mu^+$ and  $D^+X\mu^+$ combinations that pass the selections. These yields provide good estimates of the fake muon contributions in the signal samples, which are very small. Following the procedure in Ref.~\cite{LHCb-PAPER-2018-050}, we find the cross-feed corrections  of $\Bsb\to (D^0+D^+)X\mu^-\overline{\nu}$ and $\Lb\to (D^0+D^+)X\mu^-\overline{\nu}$ to be twice the measured yields for $\Bsb\to D^0 K^+X\mu^-\overline{\nu}$, which are $8500\pm340$ (7\tev) and $69~\!390\pm1130$ (13\tev), and for $\Lb\to D^0 pX\mu^-\overline{\nu}$, which are $2330\pm 140$ (7\tev) and $33~\!050\pm 460$  (13\tev).
Relative efficiencies for detecting final states with a single extra hadron are taken into account when subtracting these yields.

\subsection{\boldmath Efficiencies for $B\to D^0X\mu^-\overline{\nu}$ and  $B\to D^+X\mu^-\overline{\nu}$ }
\begin{figure}[t]
  \begin{center}
    \includegraphics[width=0.40\linewidth]{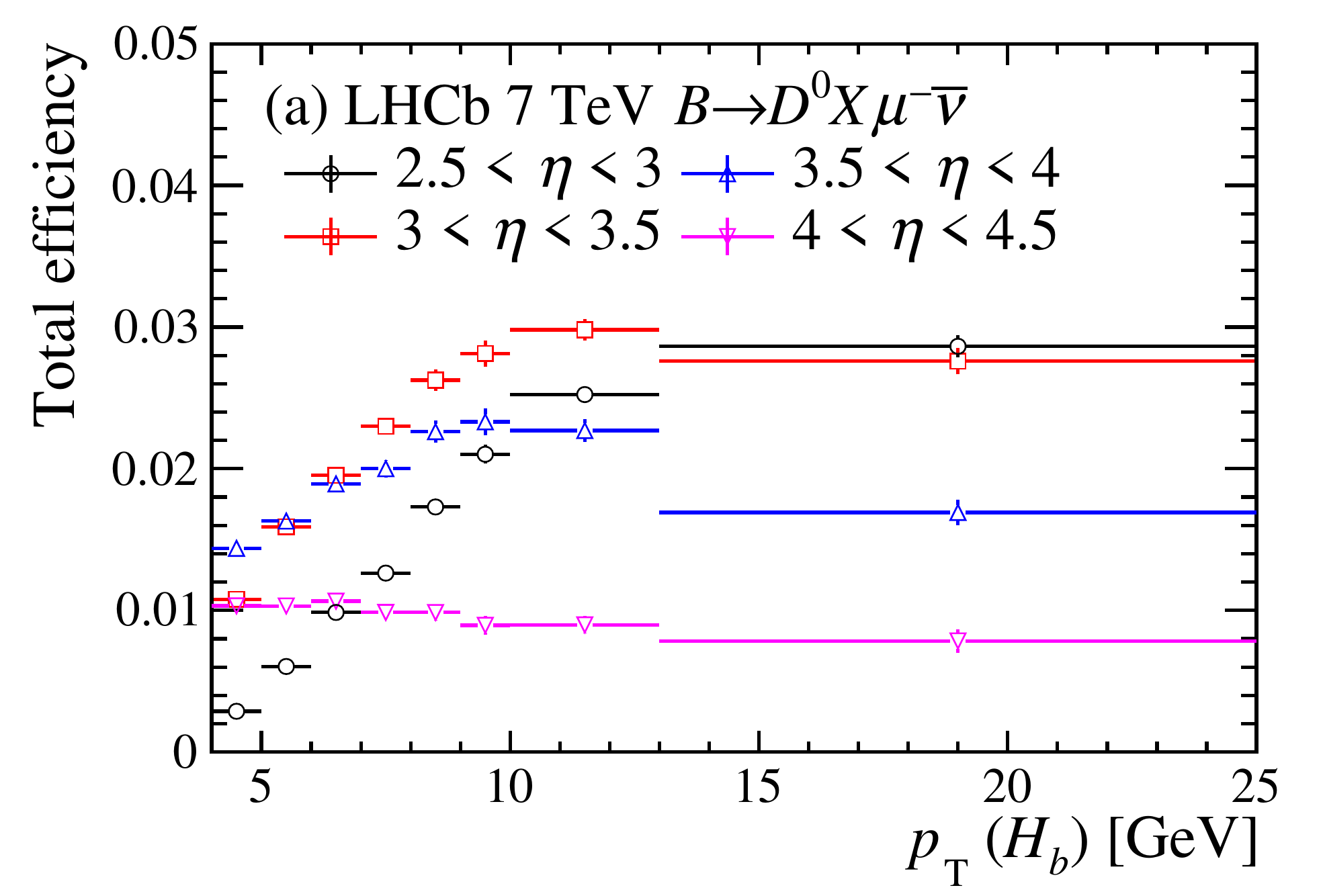}
    \includegraphics[width=0.40\linewidth]{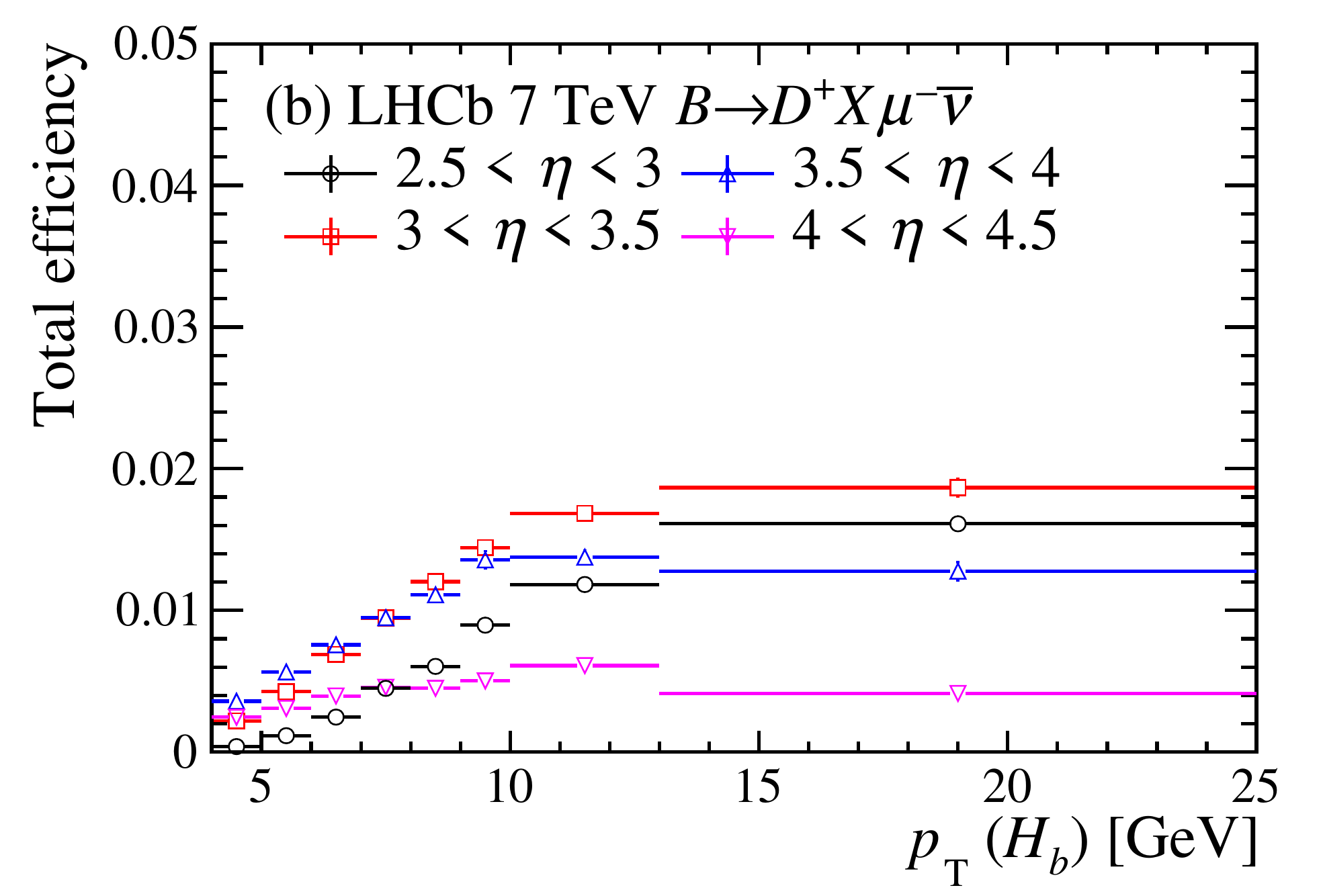}\\
    \includegraphics[width=0.40\linewidth]{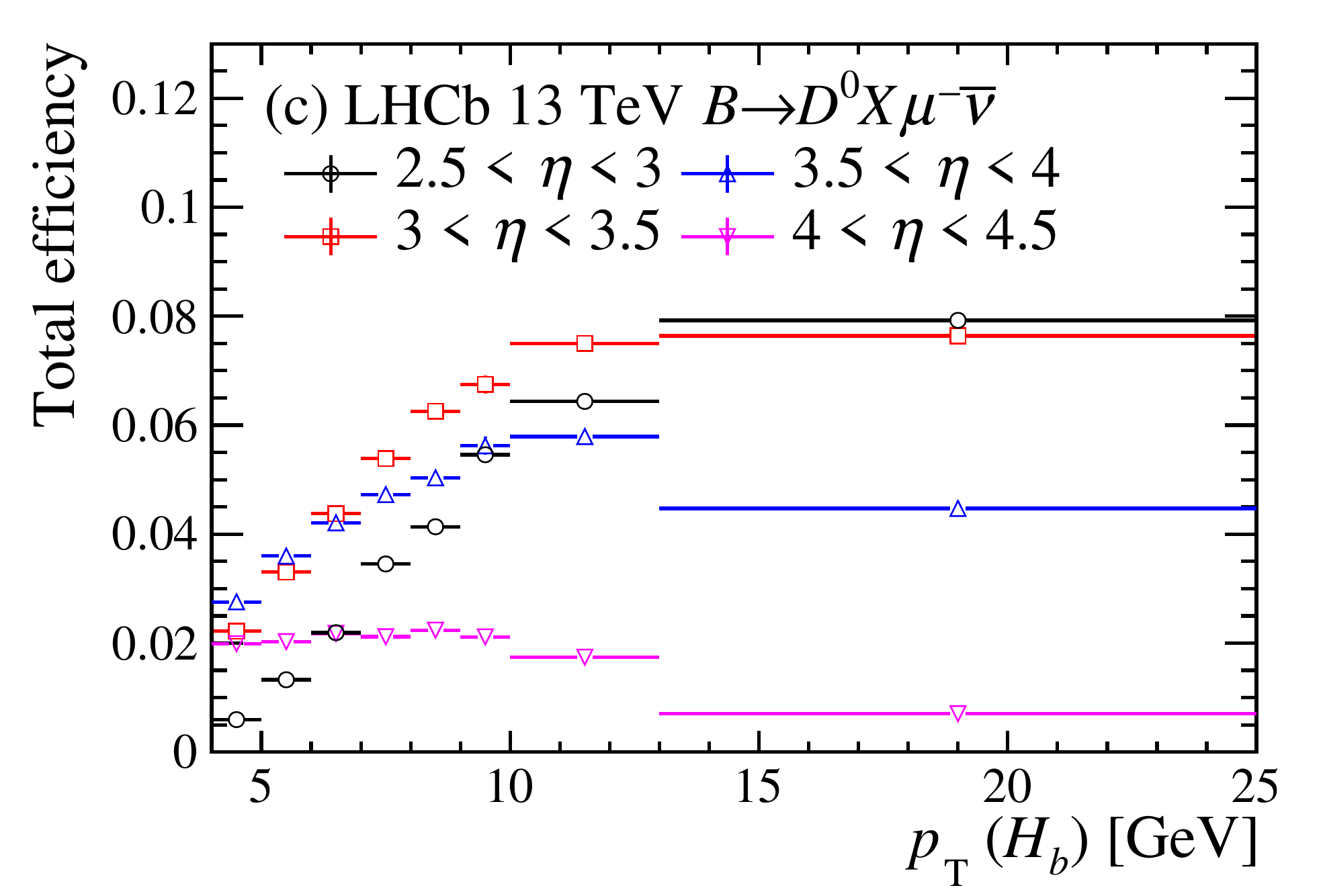}
    \includegraphics[width=0.40\linewidth]{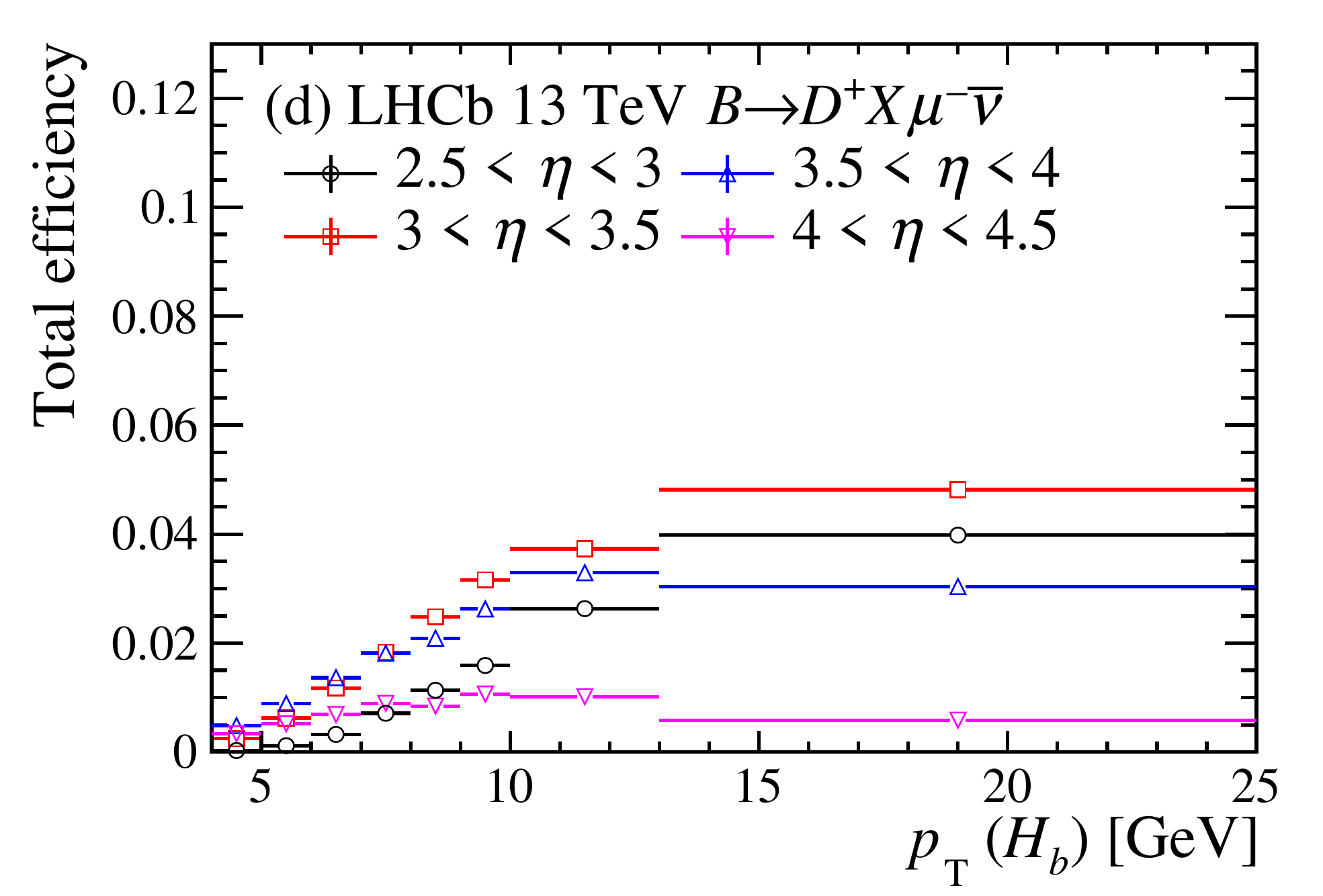}
    \vspace*{-0.5cm}
  \end{center}
  \caption{
     Total efficiencies for the  (a) $\Dz X \mu^-\overline{\nu}$ and (b) $\Dp X \mu^-\overline{\nu}$ signals in 7\tev and (c) and (d) in 13\tev samples as functions of \pt in $\eta$ intervals.
  }
  \label{fig:efficiency_total_Bsl}
\end{figure}

Similar methods based on data, as implemented for the \Bcm decay, are used to evaluate the efficiencies for trigger and particle identification. Simulation is also used to determine the efficiencies of event selection and reconstruction of these modes. The total efficiencies for $B$ meson decays into $D^0X\mu^-\overline{\nu}$ and $D^+X\mu^-\overline{\nu}$  are shown in Fig.~\ref{fig:efficiency_total_Bsl}.

\section{Results}
\subsection{\boldmath Corrections to the $\pt(H_b)$ distributions due to the missing neutrino}
\begin{figure}[b]
  \begin{center}
    \includegraphics[width=0.45\linewidth]{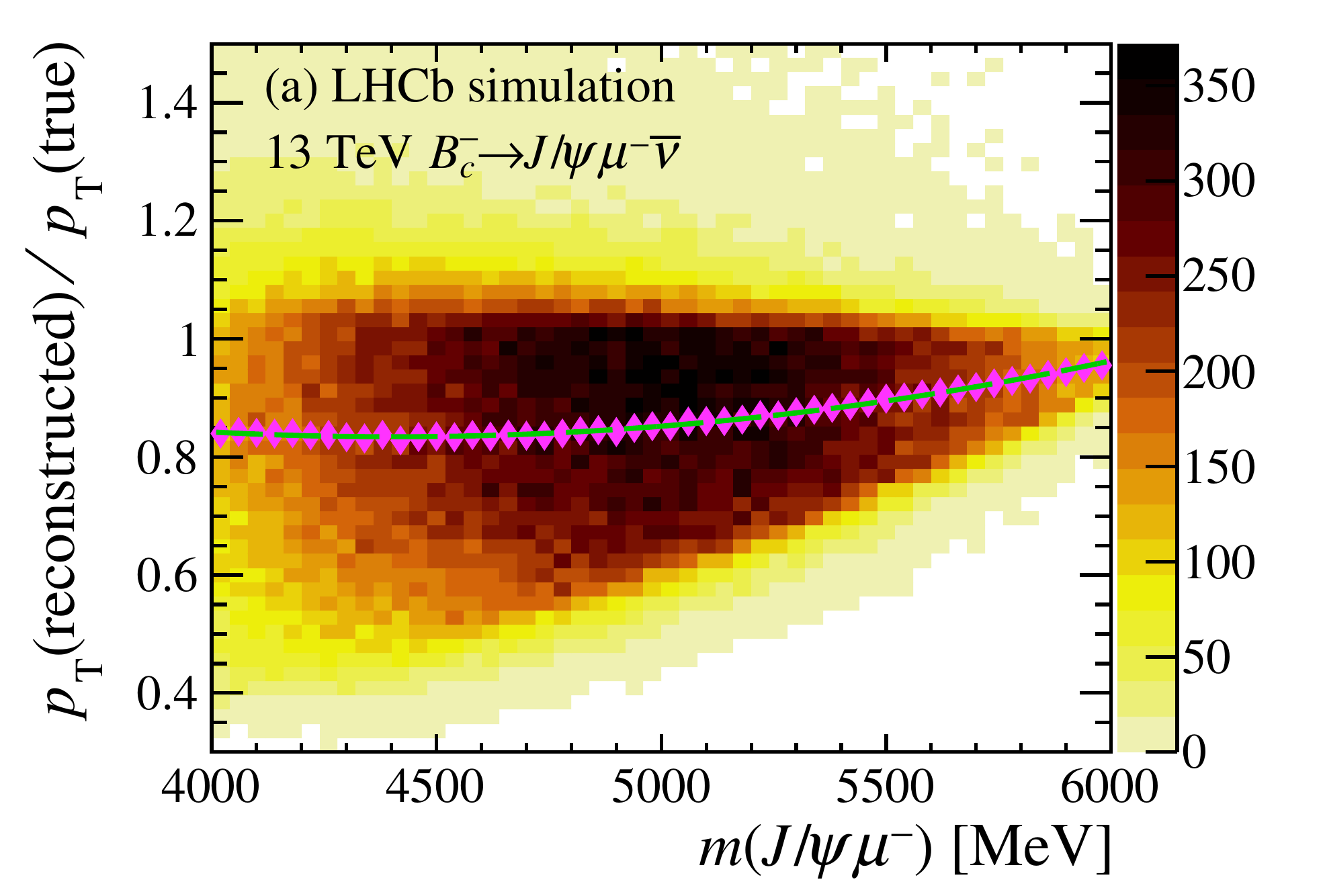}
    \includegraphics[width=0.45\linewidth]{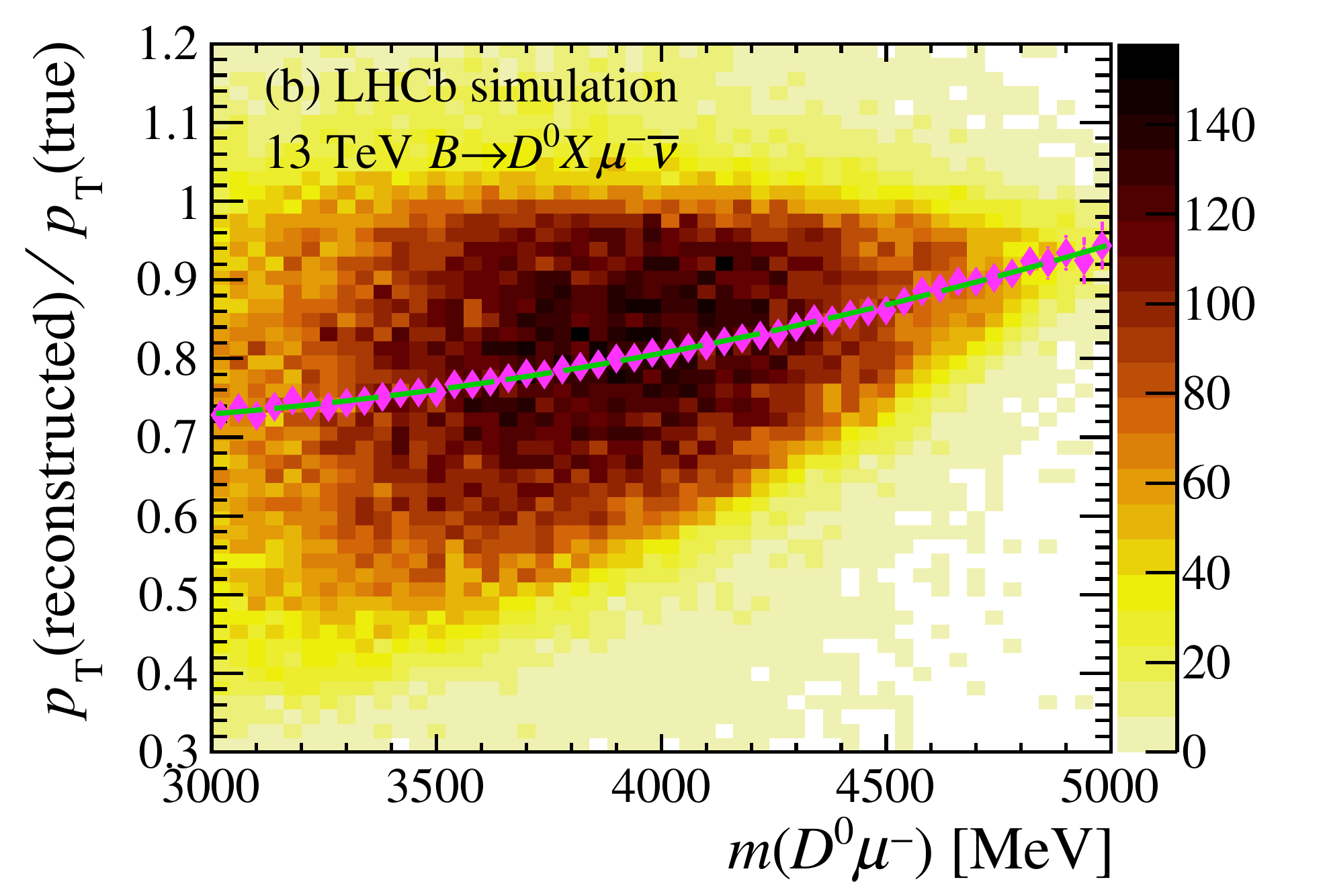}
    \includegraphics[width=0.45\linewidth]{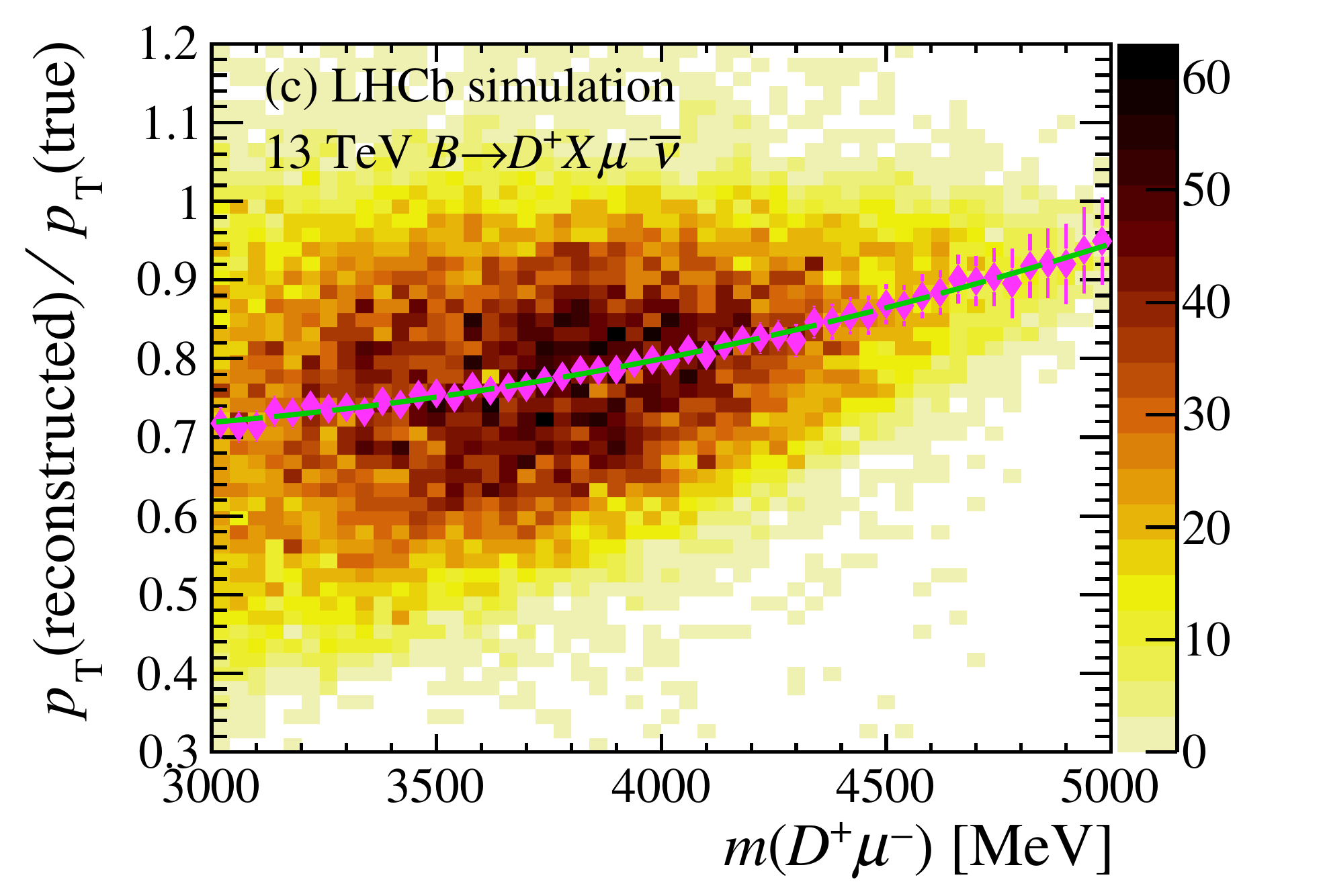}
    \vspace*{-0.5cm}
  \end{center}
  \caption{
   The  $k$-factor corrections as a function of invariant mass of (a) $m(\jpsi\mu^-)$, (b) $m(\Dz\mu^-)$, and (c) $m(\Dp\mu^-)$ for the 13\tev simulation samples. (The 7\tev results are almost identical.) The points (magenta) are the average $k$-factor corrections, and the (green) dashed line shows a second-order polynomial fit to the average data.
  }
  \label{fig:kFactorVSInvMass}
\end{figure}

Since the production kinematics of $B$ and \Bcm mesons can differ as functions of $\pt(H_b)$ and $\eta$, we need to measure $f_c/(f_u+f_d)$ as functions of these variables. The measurement of $\eta$ is straightforward, however, we do not measure directly the $\pt(H_b)$ of the $b$-flavored hadron because of the missing neutrino, and in the case of the $B$ meson possible missing extra particles. Following a procedure similar to the one used in Ref.~\cite{LHCb-PAPER-2018-050}, we determine a correction factor, $k$, that is the ratio of the average reconstructed to true $\pt(H_b)$ as a function of the invariant mass of the charmed hadron plus muon. The ratio distribution as a function of hadron-muon invariant mass are shown in Fig.~\ref{fig:kFactorVSInvMass}. The average correction, the $k$--factor, is shown on the figure. For the $B$ meson it varies from 0.75 to unity over the interval from 3~GeV to the $B$ mass, and  for the \Bcm meson it varies from 0.85 to unity over the interval from 4~GeV to the \Bcm mass. 

\subsection{\boldmath \Bcm fraction results}
\begin{figure}[b]
  \begin{center}
    \includegraphics[width=0.45\linewidth]{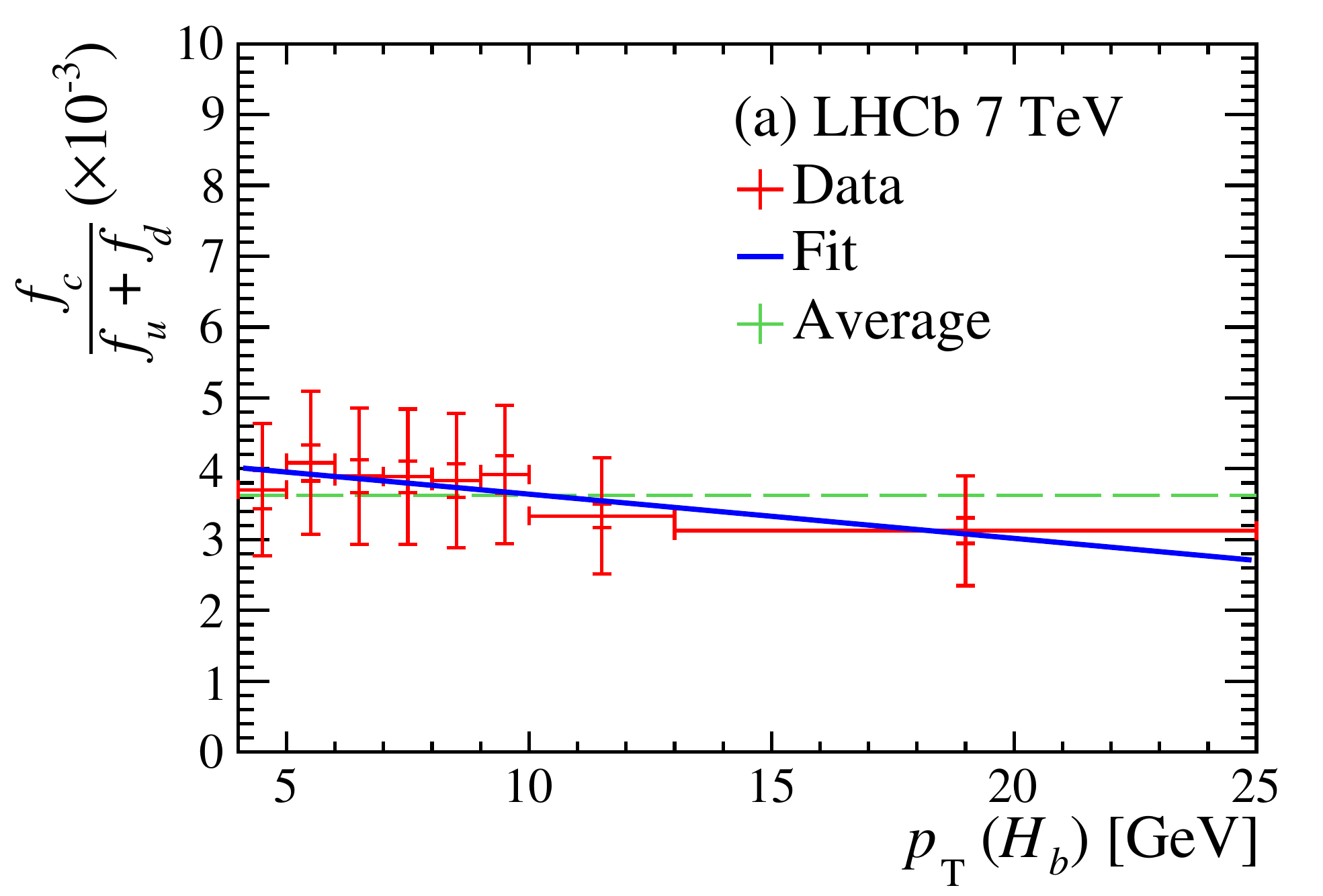}
    \includegraphics[width=0.45\linewidth]{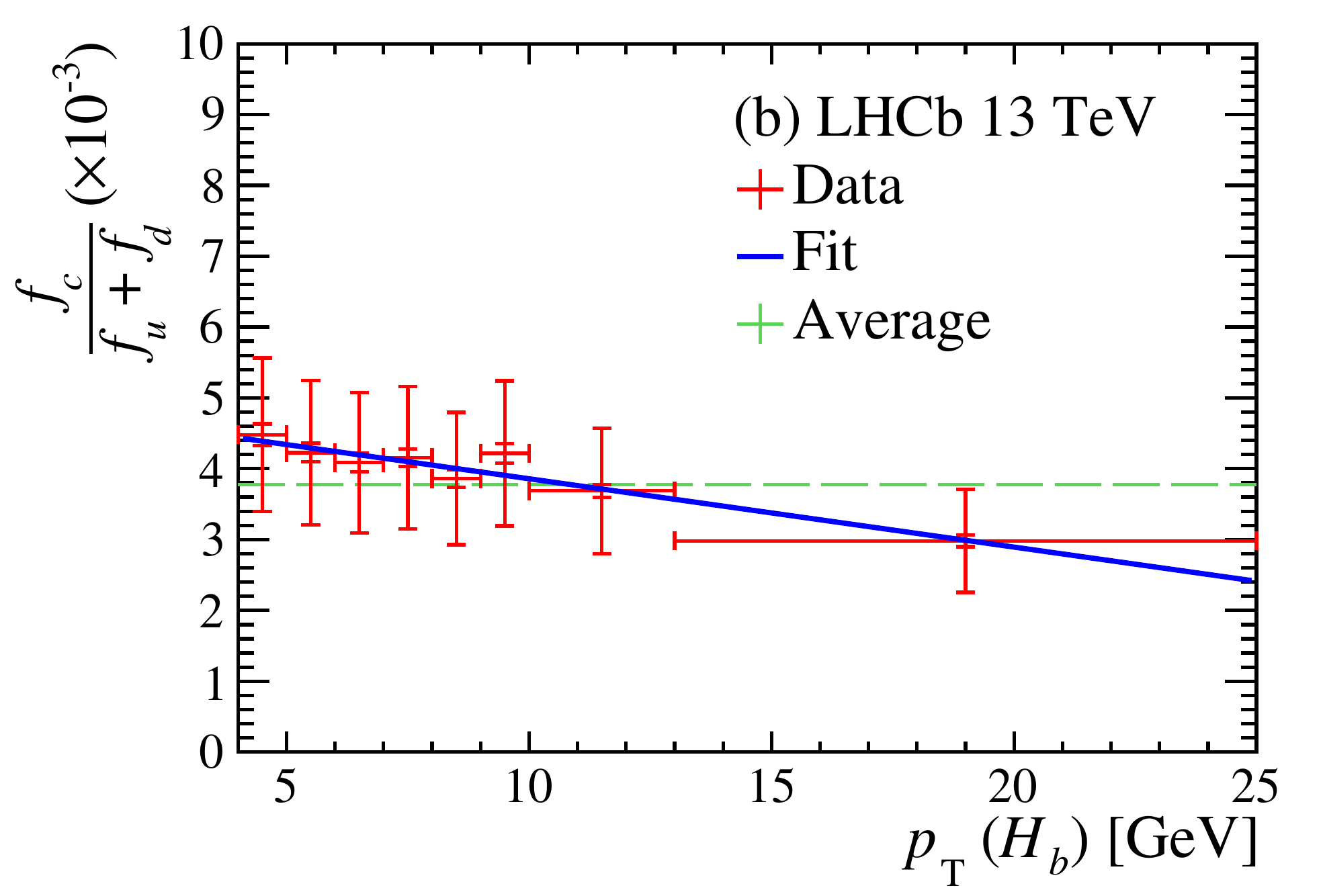}\\
    \includegraphics[width=0.45\linewidth]{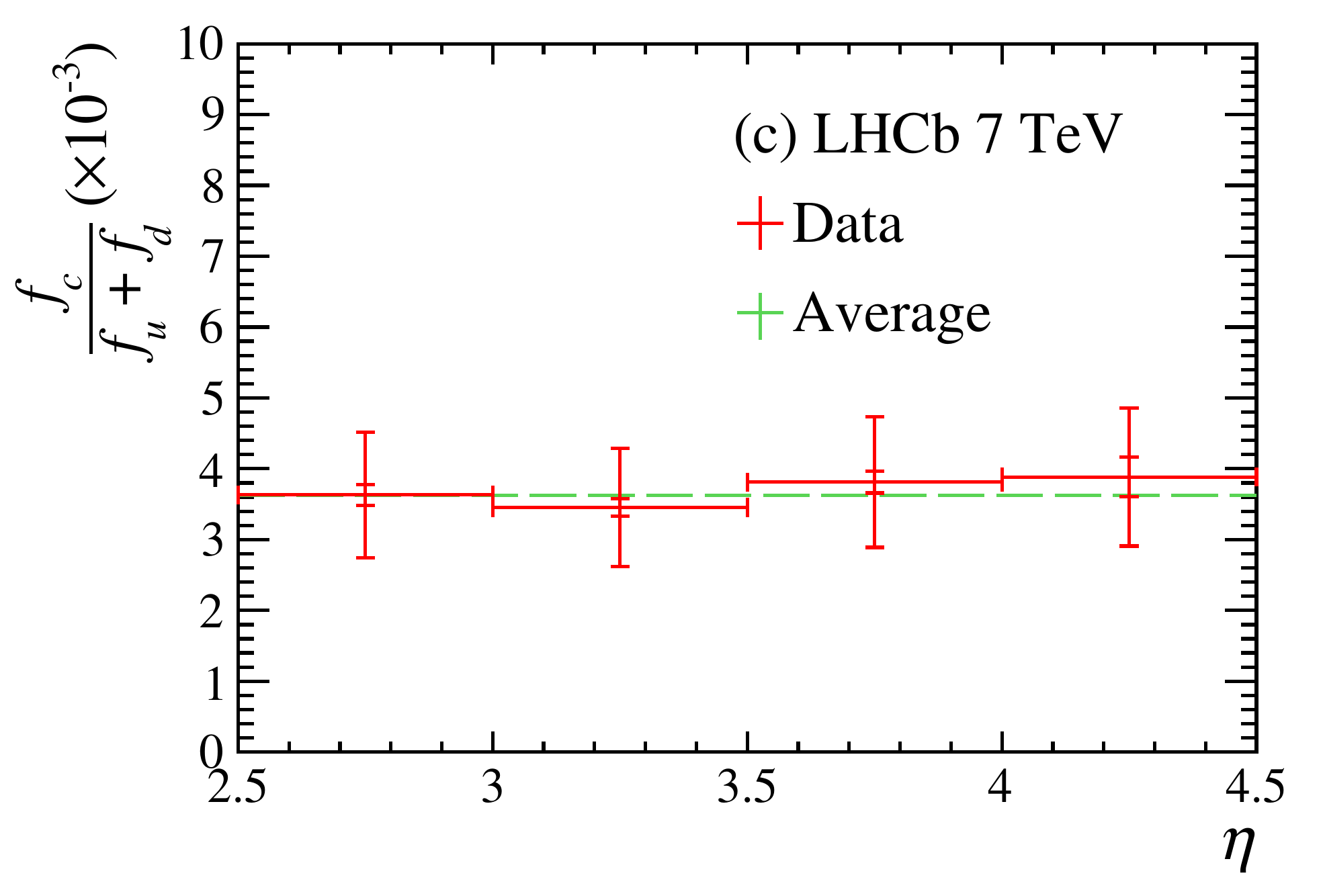}
    \includegraphics[width=0.45\linewidth]{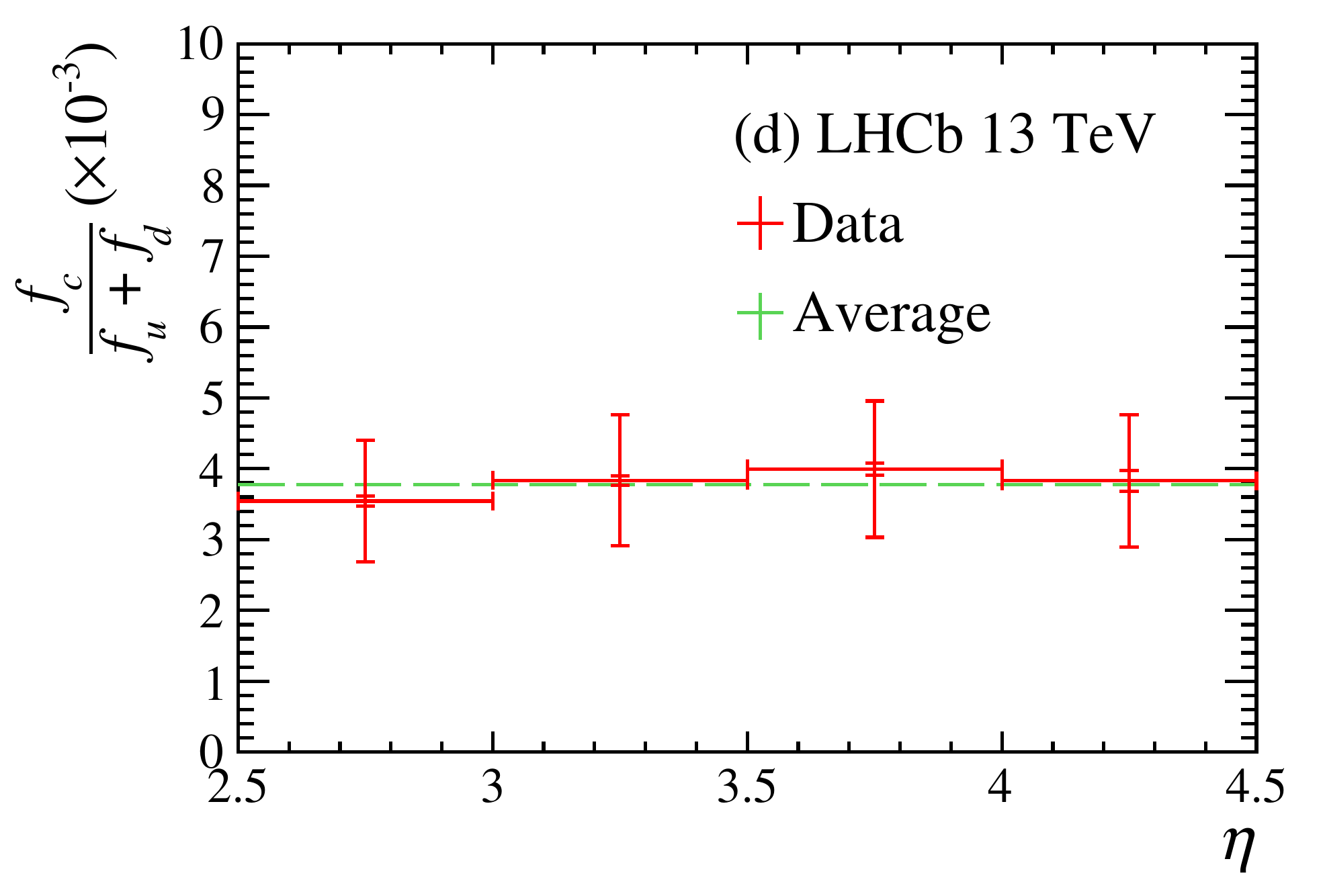}\\
    \vspace*{-0.5cm}
  \end{center}
  \caption{
     Ratio of production fractions after the $k$-factor correction as a function of (a) $\pt(H_b)$ and
     (c) $\eta$ in 7\tev data and (b) and (d) in 13\tev data. The smaller error bars show the statistical uncertainties
     and the larger ones include the statistical and systematic uncertainties. The horizontal (green) dashed-lines show the average values.
  }
  \label{fig:results_BcProdFraction_1D}
\end{figure}

The ratio of production fractions, $f_c/(f_u+f_d)$, are shown as functions of $\pt(H_b)$ and $\eta$ in Fig.~\ref{fig:results_BcProdFraction_1D}. There is little dependence on $\eta$, but the decrease as a function of $\pt(H_b)$  is noticeable.

To describe the $\pt(H_b)$ dependence we use an equation of the form
\begin{equation}
\label{eq:ptfit}
\frac{f_c}{f_u+f_d}(\pt)= A\left[p_1+p_2\left(\pt(H_b)-\langle\pt\rangle\right)\right],
\end{equation}
where $A$ represents the overall normalization and contains the total global systematic uncertainty,  thus, $A=1\pm 0.24$;\footnote{See Section~\ref{sec:systematic} for the discussion of the systematic uncertainties.} $\langle\pt\rangle$ is taken as 7.2~\!GeV, close to the average \pt of the \Bcm. The slopes, $p_2$, are similar in size to those measured for the $B_s$ meson fraction ratio as a function of \pt \cite{LHCb-PAPER-2018-050, LHCb-PAPER-2012-037}. 
Results of fits to the data using Eq.~\ref{eq:ptfit} are listed in Table~\ref{tab:results-pt}.
\begin{table}[t]
\caption{Results of the fits to Eq.~\ref{eq:ptfit}}.
\begin{center}
   \begin{tabular}{rcc}
      \hline\hline
      Energy &$p_1$ & $p_2\cdot 10^{-2}$ (GeV$^{-1}$)\\\hline
 7\tev & $3.82\pm 0.09\pm 0.05$ & $-6.2\pm 1.7\pm 1.1$      \\
 13\tev & $4.13\pm 0.05\pm 0.04$ & $-9.7\pm 0.8\pm 1.0$ \\     
       \hline\hline
   \end{tabular}\end{center}
   \label{tab:results-pt}
\end{table}

The average fractions in the interval $4<\pt(H_b)<25$\gev are found by integrating over $\pt(H_b)$.
To allow for facile changes to our results due to improved theoretical predictions,  we provide the results for
\begin{align*}
\frac{f_c}{f_u+f_d}\cdot {\cal{B}}(\Bcm\to\jpsi\mu^-\overline{\nu})=&\left(7.07\pm 0.15\pm 0.24\right)\cdot 10^{-5} {\rm ~for ~7\tev},\\\nonumber
\frac{f_c}{f_u+f_d}\cdot {\cal{B}}(\Bcm\to\jpsi\mu^-\overline{\nu})=&\left(7.36\pm 0.08\pm 0.30\right)\cdot 10^{-5} {\rm ~for ~13\tev}.\\\nonumber
\end{align*}

Next we give the result on the fractions ratio
\begin{align*}
\frac{f_c}{f_u+f_d}=&\left(3.63\pm 0.08\pm 0.12\pm 0.86\right)\cdot 10^{-3} {\rm ~for ~7\tev},\\\nonumber
\frac{f_c}{f_u+f_d}=&\left(3.78\pm 0.04\pm 0.15\pm 0.89\right)\cdot 10^{-3} {\rm ~for ~13\tev},\nonumber
\end{align*}
where the third uncertainty is due to the theoretical prediction of ${\cal{B}}(\Bcm\to\jpsi\mu^-\overline{\nu}).$ To find $f_c/f_u$ just double these numbers.

\begin{figure}[b]
  \begin{center}
    \includegraphics[width=0.45\linewidth]{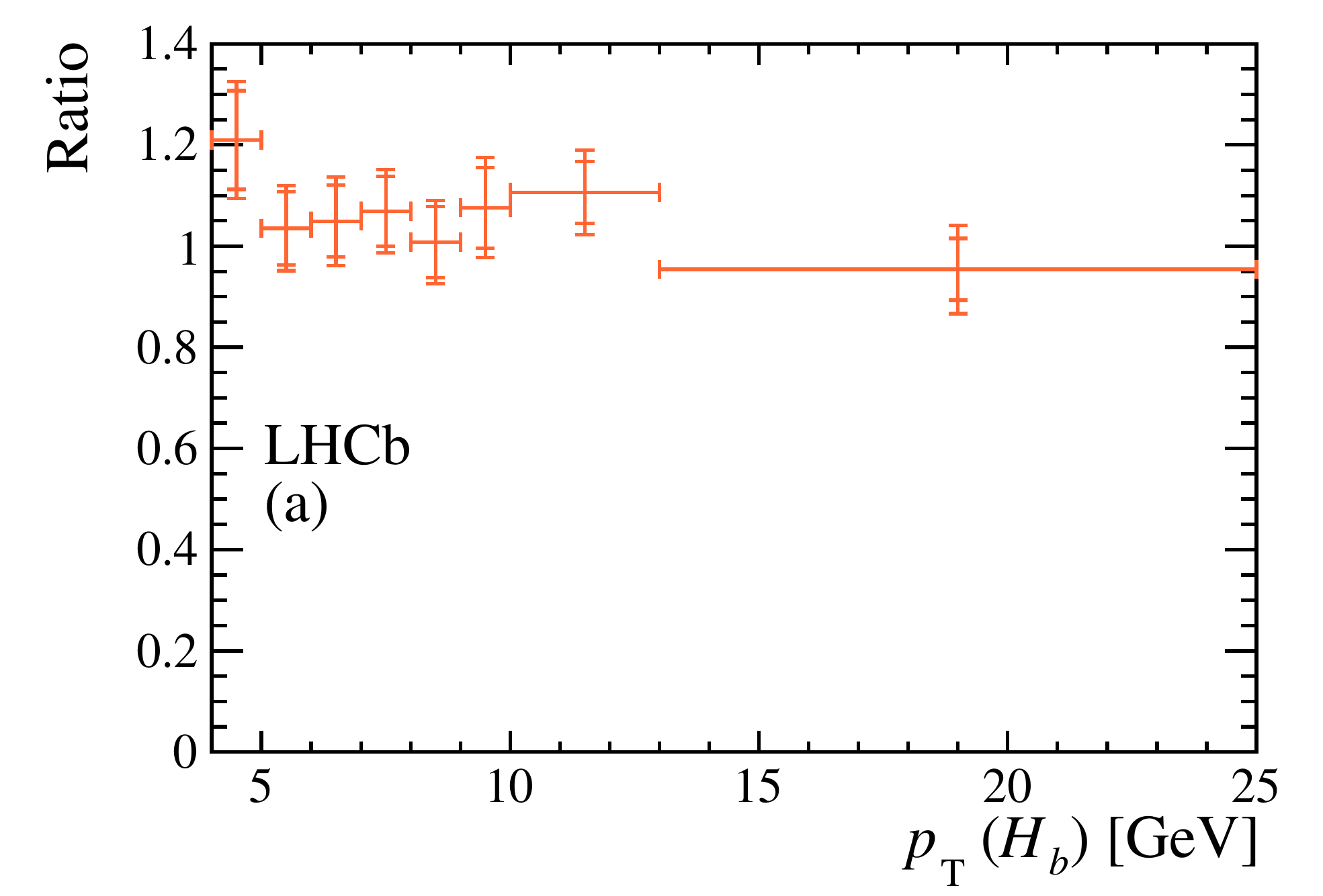}
    \includegraphics[width=0.45\linewidth]{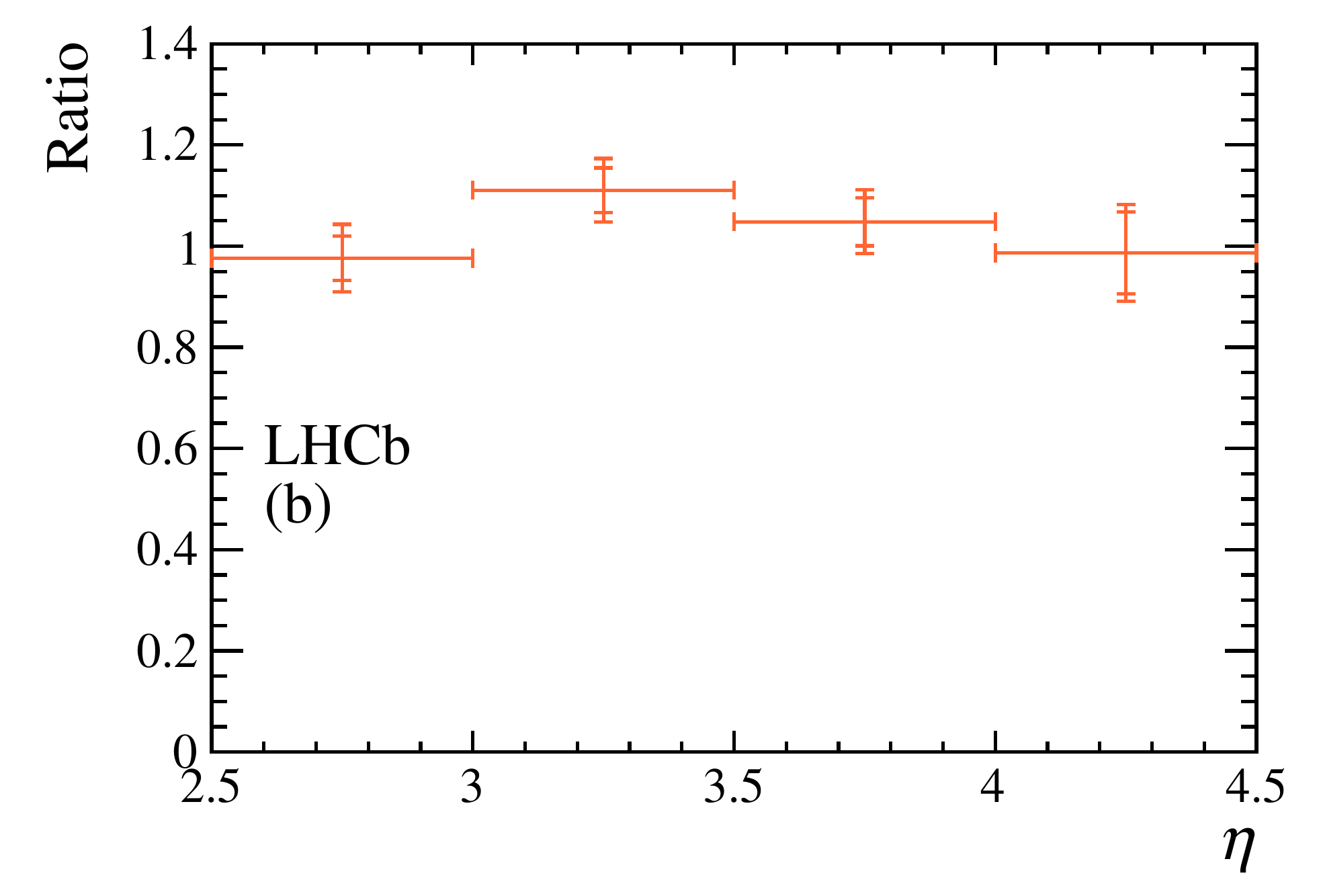}
    \vspace*{-0.5cm}
  \end{center}
  \caption{
     Ratio of the $\Bcm$ production fractions at 13 TeV to 7 TeV as a function of (a) $\pt(H_b)$ and
     (b) $\eta$. The smaller error bars show the statistical uncertainties
     and the larger ones include the statistical and systematic uncertainties added in quadrature. 
  }
  \label{fig:results_BcProdFractionRatio_1D}
\end{figure}

We also measure the ratio of the $\Bcm$ production fraction at $7\tev$ to that at $13\tev$. Figure~\ref{fig:results_BcProdFractionRatio_1D} shows 
the ratio  as functions of \pt and $\eta$. Here most of the systematic uncertainties cancel. The
integrated value of the ratio of $13\tev$ and $7\tev$ is measured as $1.02\pm 0.02\pm 0.04$, consistent with no increase in the \Bcm fraction ratio as a function of center-of-mass energy.

The \Bcm fraction with respect to inclusive $b$--hadron production can be derived from the information in previous LHCb $b$--hadron fraction papers Ref.~\cite{LHCb-PAPER-2018-050,LHCb-PAPER-2012-037,LHCb-PAPER-2011-018}.
There the measured values of the ratios of $b$--hadron fractions over the same \pt range in terms of the $b$--hadron \pt are for \Bsb mesons ($f_s$) and \Lb baryons 
\begin{align}
\frac{f_s}{f_u+f_d}&=\left\{
\begin{aligned}
&0.124\pm0.010 & & ~~\!(7\tev)\text{\cite{LHCb-PAPER-2012-037}}\\
&0.122\pm0.006 & & (13\tev)\text{\cite{{LHCb-PAPER-2018-050}}},\\
\end{aligned}\right.\\
\frac{f_{\Lb}}{f_u+f_d}&=\left\{
\begin{aligned}
   &0.223\pm0.036 & & ~~\!(7\tev)\text{\cite{LHCb-PAPER-2011-018}}\\
&0.259\pm0.018 & & (13\tev)\text{\cite{{LHCb-PAPER-2018-050}}},\\
\end{aligned}\right.
\end{align}
where the uncertainties contain both statistical and systematic components added in quadrature.  For the measurement of the
$f_\Lb$ fraction at 7\tev, the dominant systematic uncertainty is from the lack of the knowledge of $\BR(\Lc\to\proton\Km\pip)$ at that time \cite{LHCb-PAPER-2011-018}; here the value and uncertainty have been recalculated according to the latest value of $\BR(\Lc\to\proton\Km\pip)$ from the PDG \cite{PDG2018}.

Taking the sum of all the $b$-hadron fractions to be unity, and ignoring $f_c$ here because it is so small,
\begin{equation}
f_u+f_d+f_s+f_{\Lb}(1+\delta)=1,
\end{equation}
where $\delta=0.25\pm0.10$ is a correction factor derived in Ref.~\cite{LHCb-PAPER-2016-031} that accounts for heavier $b$--baryons, mainly the $\PXi_b$. Solving for $f_u+f_d$ yields
\begin{align}\label{eqn:budfraction_fud}
f_u+f_d &= \left(1+\frac{f_s}{f_u+f_d}+\frac{f_{\Lb}}{f_u+f_d}(1+\delta)\right)^{-1},\nonumber\\
             &=\left\{
\begin{aligned}
&0.713\pm0.026 & & ~~\!(7\tev)\\
&0.692\pm0.015 & & (13\tev).\\
\end{aligned}\right.
\end{align}
We find that   
\begin{equation}\nonumber
\label{eqn:budfraction_Allb-Br}
   {f_c}\cdot {\cal{B}}(\Bcm\to\jpsi\mu^-\overline{\nu})=\left\{
	\begin{aligned}
	   &(5.04\pm0.11\pm0.17\pm 0.18)\cdot 10^{-5}&(7\tev)\\
	   &(5.09\pm0.06\pm0.21\pm0.11)\cdot 10^{-5}&(13\tev)\\
	\end{aligned}\right.,
\end{equation}
where the first uncertainty is statistical, the second is systematic, and the third is 
from the fractions of the $\Bs$ and $\Lb$ given in Eq.~\ref{eqn:budfraction_fud}. 
We also provide the result for $f_c$,  

\begin{equation}\nonumber
\label{eqn:budfraction_Allb}
   {f_c}=\left\{
	\begin{aligned}
	   &(2.58\pm0.05\pm0.62\pm0.09)\cdot 10^{-3}&(7\tev)\\
	   &(2.61\pm0.03\pm0.62\pm0.06)\cdot 10^{-3}&(13\tev)\\
	\end{aligned}\right.,
\end{equation}
where the first uncertainty is statistical, the second is systematic
including that from $\BR(\Bcmu)$ and the third is 
from the fractions of the $\Bs$ and $\Lb$ given in Eq.~\ref{eqn:budfraction_fud}.  

\subsection{\boldmath The $\Bcm-\Bc$ production asymmetry}
The production asymmetries are measured in two different magnetic field configurations and then averaged. No significant asymmetry is observed in any intervals of $\pt(H_b)$ or $\eta$. The results are summarized in 
Table~\ref{tab:results_BcProdAsymmetry_2D}.
\begin{table}[h]
   \caption{
	The $\Bcm-\Bc$ production asymmetry ($\times10^{-2}$) as a function of
	$\pt(H_b)$ and $\eta$ at 7 TeV and 13 TeV. 
   }
\begin{center}
   \begin{tabular}{rll}
	\hline\hline
	\multicolumn{3}{c}{7 TeV production asymmetry}\\
	\hline
	$\pt~\!{\rm (GeV)}\backslash\eta$ &	 ~~~~~~$2.5-3.5$&	~~~~~~~~$3.5-4.5$\\
$4-6$~~ &~~~\!$7.91\pm7.00\pm1.03$	& 	~~\!$-6.44\pm6.44\pm2.10$	\\
$6-8$~~ & 	$-4.34\pm5.43\pm1.62$	& 	~~\!$-6.66\pm6.65\pm2.03$	\\
$8-10$~~ 	 &	$-1.13\pm6.31\pm1.56$	& 	~~\!$-9.63\pm7.23\pm0.81$	\\
$10-25$~~  &	~~~\!$0.24\pm4.13\pm0.98$	& 	~~\!$-4.87\pm8.63\pm1.44$	\\
	\hline
	\multicolumn{3}{c}{13 TeV production asymmetry}\\
	\hline
$\pt~\!{\rm (GeV)}\backslash\eta$ &	 ~~~~~~~$2.5-3.5$&	~~~~~~~~$3.5-4.5$\\
$4-6$~~  &	~~~$3.13\pm3.33\pm1.16$	& 	~~~~\!$1.76\pm3.23\pm0.91$	\\
$6-8$~~  &	~\!$-0.34\pm2.79\pm1.26$	& 	~$-5.03\pm3.61\pm1.06$	\\
$8-10$~~ &	~~~$2.03\pm2.73\pm0.94$	& 	~$-2.48\pm4.29\pm1.78$	\\
$10-25$~~  &	~~~$1.50\pm2.05\pm0.73$ & 	~$-1.47\pm4.20\pm2.18$	\\	\hline
  \hline
  \end{tabular}
\end{center}\label{tab:results_BcProdAsymmetry_2D}
\end{table}

Averaging the $\Bcm-\Bc$ production asymmetries over $\pt(H_b)$ and $\eta$, we find \mbox{$(-2.5\pm 2.1\pm 0.5)\%$}, and $(-0.5\pm 1.1\pm 0.4)\%$ at center-of-mass energies of 7 and 13\tev, respectively.

\section{Systematic uncertainties}
\label{sec:systematic}
Systematic uncertainties are separated into two categories: ``global", which apply across the phase space, and ``local", which are calculated in each two-dimensional $\pt(H_b)-\eta$ bin. These uncertainties are listed in Table~\ref{tab:systematic}.
\begin{table}[htb]
   \caption{Summary of the relative systematic uncertainties for $f_c/(f_u+f_d)$(\%) and the absolute production asymmetries $a_{\rm prod}$(\%). For local uncertainties, the ranges correspond to the minimum and maximum uncertainties evaluated in the $\pt(H_b)$ and $\eta$ ranges.
   } 
\begin{center}
   \begin{tabular}{lcccc}
      \hline\hline
	& \multicolumn{2}{c}{$f_c/(f_u+f_d)$} & \multicolumn{2}{c}{$a_{\rm prod}$}\\
	\hline
	& 7 \tev & 13 \tev & 7 \tev & 13\tev \\
 {\it Local  uncertainties} & & & &\\
  Signal shape  & 0.12--9.56&0.14--2.80 & 0.04--1.80& 0.01--0.78\\
  Background shape & 0.34--6.16& 0.02--5.80& 0.06--3.05& 0.05--2.45\\
  Feed-down channels & 0.12--5.00& 0.43--2.27& 0.01--1.11 & 0.03--0.65\\
  Decay models & 0.00--2.00& 0.01--3.84& 0.02--0.28&0.02--0.61\\
  Muon ID in $\jpsi\mu^-$ & 0.06--5.79&0.03--2.92& 0.02--0.37& 0.01--0.18\\
  Trigger for $\jpsi$ & 0.00--0.23 & 0.00--0.34& 0.05--2.34& 0.07--4.24\\
  Simulation decay model& 0.00--2.00 & 0.01--3.84&0.02--0.28& 0.02-0.61 \\
  Hadron ID in $DX\mu^-\overline{\nu}$ & 0.04--1.81 & 0.01--2.01 & --&--\\
  Muon trigger \& ID in $DX\mu^-\overline{\nu}$& 0.02--1.34& 0.00--0.21&--&--\\
  Simulation sample size& 1.5--11.5& 2.1--10.7&0.5--1.1&0.5--1.2\\
  $k$-factor& 0.02--0.95 & 0.05--0.70& 0.01--0.10& 0.00--0.10\\
  Tracking asymmetry & --&--& 0.00--0.28&  0.00--0.09\\
   {\it Global  uncertainties} & & & &\\
$\BR(\jpsi\to\mumu)$ & $0.55$ & $0.55$ & $-$ & $-$\\
$\BR(D^+\to K^-\pi^+\pi^+)$ or $\BR(D^0\to K^-\pi^+)$ & $1.0$ & $1.0$ & $-$ & $-$\\
	$\BR(B\to H_c X\mu^-\overline{\nu})$ & $1.8$ & $1.8$ & $-$ & $-$\\
	 Cross-feed contribution & $0.2$ & $0.2$ & $-$ & $-$\\
	 Multiplicity cut & 1.2&2.7 & $-$ & $-$\\
	 Tracking efficiency & $1.8$ & $1.8$ & $-$ & $-$\\
	\hline
{\it Uncertainty sum}  & 4.3--21.3& 5.1--17.4 & 1.0--3.5 & 1.0--4.8\\
 $\BR(\Bcmu)$ & 23.6 & 23.6 &--&--\\
  {\it Overall uncertainty}& 24.0--31.8& 24.1--29.3 & 1.0--3.5 & 1.0--4.8\\
	 \hline\hline
	 \end{tabular}
\end{center}\label{tab:systematic}
\end{table}

First let us consider the $\Bcmu$ decay. The uncertainty due to the signal shape used to fit the $m_{\rm cor}$ distribution is determined by changing the baseline signal shape, the sum of a double sided Crystal Ball function and a bifurcated Gaussian, to a kernel estimation. 
To find the shape of the combinatorial and misidentification backgrounds we use simulated inclusive samples of $b\to\jpsi X$ events not including \Bcm decays. A total of 500 samples are generated and different fits to the samples are performed to determine the possible uncertainty. This procedure is also used for the $a_{\rm prod}$ measurement. We call contributions to the $\jpsi\mu^-$ mass spectrum ``feed-down" contributions,  occurring from other \Bcm decay channels including $\jpsi\tau\overline{\nu}$, $\psi(2S)\mu^-\overline{\nu}$, and $\chi_c\mu^-\overline{\nu}$. The systematic uncertainty results from the uncertainties in their branching fractions. Different decay models for \Bcmu decays can change the $m_{\rm cor}$ shape. We use the model of Ebert \etal  \cite{Ebert:2010zu} for our baseline prediction. Then we also use the model by Kiselev \cite{Kiselev:2002vz} to find the efficiencies and take half the difference as the systematic uncertainty.  We also estimate the uncertainty due to the sensitivity to various selection requirements and simulation statistics. The muon identification efficiencies are determined from data using inclusive samples of \jpsi decay where one of the muon candidates is not identified. The trigger efficiency is determined by using three independent samples of events, those that trigger on a $\jpsi$, those that triggered on something else in the event, and those that trigger on both the \jpsi and something else. These samples are then used to compute the trigger efficiencies in two-dimensional $\pt(H_b)$ and $\eta$ bins.

Next, we turn to the $B\to DX\mu\nu$ modes. The efficiencies and their uncertainties for identifying pions and kaons are determined by using almost background free samples of $D^{*+}\to \pi^+D^0,~D^0\to K^-\pi^+$ decays. The trigger  and muon identification efficiencies, and their uncertainties, are obtained in the same manner as for the $\Bcmu$ mode. There are small systematic uncertainties related to efficiency estimates and the assumed $D^*$ to $D$ mixtures, as well as simulation statistics. 
Global systematic uncertainties include the hadron branching fractions listed in Table~\ref{tab:charmdecay}, cross-feed corrections arising from \Bsb and \Lb decays into $D X\mu^-\overline{\nu}$ events, and a global hadron plus photon multiplicity requirement. The latter is evaluated with data. 

\section{Conclusions}
In 7 and 13\tev $pp$ collisions the product of ${\cal{B}}(\Bcmu)$ with the relative fraction of \Bcm mesons with respect to the sum of \Bz and \Bp mesons in the ranges $2.5<\eta<4.5$ and $4<\pt(H_b)<25\gev$ is found to be
\begin{align*}
\frac{f_c}{f_u+f_d}\cdot {\cal{B}}(\Bcm\to\jpsi\mu^-\overline{\nu})=&\left(7.07\pm 0.15\pm 0.24\right)\cdot 10^{-5} {\rm ~for ~7\tev},\\\nonumber
\frac{f_c}{f_u+f_d}\cdot {\cal{B}}(\Bcm\to\jpsi\mu^-\overline{\nu})=&\left(7.36\pm 0.08\pm 0.30\right)\cdot 10^{-5} {\rm ~for ~13\tev}.\\\nonumber
\end{align*}

We derive the product of  ${f_c}\cdot {\cal{B}}(\Bcm\to\jpsi\mu^-\overline{\nu})$ at the two energies as
\begin{equation}\nonumber
\label{eqn:budfraction_Allb-Br2}
   {f_c}\cdot {\cal{B}}(\Bcm\to\jpsi\mu^-\overline{\nu})=\left\{
	\begin{aligned}
           &(5.04\pm0.11\pm0.17\pm 0.18)\cdot 10^{-5}&(7\tev)\\
	   &(5.09\pm0.06\pm0.21\pm0.11)\cdot 10^{-5}&(13\tev)\\
	\end{aligned}\right. 
\end{equation}

Using the average of the theoretical prediction ${\cal{B}}(\Bcmu) = (1.95\pm 0.46)\%$, where the uncertainty is given by the standard deviation derived from the distribution of the models, we determine
\begin{align*}
\frac{f_c}{f_u+f_d}=&\left(3.63\pm 0.08\pm 0.12\pm 0.86\right)\cdot 10^{-3} {\rm ~for ~7\tev},\\\nonumber
\frac{f_c}{f_u+f_d}=&\left(3.78\pm 0.04\pm 0.15\pm 0.89\right)\cdot 10^{-3} {\rm ~for ~13\tev},\nonumber
\end{align*}
where the first uncertainties are statistical, the second systematic, and the third due to the theoretical prediction of ${\cal{B}}(\Bcm\to\jpsi\mu^-\overline{\nu}).$ There is a small dependence on the transverse momentum of the \Bc meson, but no dependence on its pseudorapidity is observed. We also report
\begin{equation}\nonumber
\label{eqn:budfraction_Allb}
   {f_c}=\left\{
	\begin{aligned}
	   &(2.58\pm0.05\pm0.62\pm0.09)\cdot 10^{-3}&(7\tev)\\
	   &(2.61\pm0.03\pm0.62\pm0.06)\cdot 10^{-3}&(13\tev)\\
	\end{aligned}\right.,
\end{equation}
where the first uncertainty is statistical, the second is systematic
including that from $\BR(\Bcmu)$ and the third is 
from the fractions of the $\Bs$ and $\Lb$ given in Eq.~\ref{eqn:budfraction_fud}.

 The ratio of fractions, $1.02\pm 0.02\pm 0.04$, for 13\tev/7\tev is consistent with no increase in the \Bcm fraction. Furthermore, using the assumption of no \CP violation in the $\Bcmu$ decay, we find that the average asymmetry in $\Bcm - \Bc$ production is consistent with zero. The measurements are 
 $(-2.5\pm 2.1\pm 0.5)\%$, and $(-0.5\pm 1.1\pm 0.4)\%$ at center-of-mass energies of 7 and 13\tev, respectively.

These results are useful to extract absolute branching fractions for \Bcm measurements, albeit with a relatively large uncertainty. They also challenge QCD calculations to predict the measured \Bcm fractions and explain the consistency between the fractions measured at 7 and 13\tev \cite{Brambilla:2010cs, Ali:2018xfq}.

\input{acknowledgements}




 
\input{Bcfrac-prl-v5.bbl}

\newpage
\input{LHCb_Authorship_10-Sep-2019}

\end{document}

%% file: preamble.tex
\usepackage[top=1in, bottom=1.25in, left=1in, right=1in]{geometry}

\usepackage{microtype}
\usepackage{lineno}  
\usepackage{xspace} 
\usepackage{caption} 

\usepackage{graphicx}  
\usepackage{color}
\usepackage{colortbl}
\graphicspath{{./figs/}} 
\DeclareGraphicsExtensions{.pdf,.PDF,png,.PNG}

\usepackage{amsmath} 
\usepackage{amssymb}
\usepackage{amsfonts}
\usepackage{upgreek} 

\newcommand*\patchAmsMathEnvironmentForLineno[1]{%
\expandafter\let\csname old#1\expandafter\endcsname\csname #1\endcsname
\expandafter\let\csname oldend#1\expandafter\endcsname\csname
end#1\endcsname
 \renewenvironment{#1}%
   {\linenomath\csname old#1\endcsname}%
   {\csname oldend#1\endcsname\endlinenomath}%
}
\newcommand*\patchBothAmsMathEnvironmentsForLineno[1]{%
  \patchAmsMathEnvironmentForLineno{#1}%
  \patchAmsMathEnvironmentForLineno{#1*}%
}
\AtBeginDocument{%
\patchBothAmsMathEnvironmentsForLineno{equation}%
\patchBothAmsMathEnvironmentsForLineno{align}%
\patchBothAmsMathEnvironmentsForLineno{flalign}%
\patchBothAmsMathEnvironmentsForLineno{alignat}%
\patchBothAmsMathEnvironmentsForLineno{gather}%
\patchBothAmsMathEnvironmentsForLineno{multline}%
\patchBothAmsMathEnvironmentsForLineno{eqnarray}%
}

\usepackage{hyperxmp}

\usepackage[pdftex,
            pdfauthor={\paperauthors},
            pdftitle={\paperasciititle},
            pdfkeywords={\paperkeywords},
            pdfcopyright={Copyright (C) \papercopyright},
            pdflicenseurl={\paperlicenceurl}]{hyperref}

\usepackage[all]{hypcap} 

\input{lhcb-symbols-def} 

\usepackage{cite} 
\usepackage{mciteplus}

%% file: lhcb-symbols-def.tex
\usepackage{xspace} 
\usepackage{upgreek}

\def\lhcb   {\mbox{LHCb}\xspace}

\def\cms    {\mbox{CMS}\xspace}

\def\babar  {\mbox{BaBar}\xspace}
\def\belle  {\mbox{Belle}\xspace}

\def\cleo   {\mbox{CLEO}\xspace}
\def\cdf    {\mbox{CDF}\xspace}

\def\MagUp {\mbox{\em Mag\kern -0.05em Up}\xspace}

\ifthenelse{\boolean{uprightparticles}}%
{

 \def\Pmu         {\ensuremath{\upmu}\xspace}                 
 \def\Pnu         {\ensuremath{\upnu}\xspace}                 
                  
 \def\Ppi         {\ensuremath{\uppi}\xspace}

 \def\Ptau        {\ensuremath{\uptau}\xspace}

 \def\Ppsi        {\ensuremath{\uppsi}\xspace}

 \def\PDelta      {\ensuremath{\Delta}\xspace}                 
 \def\PXi         {\ensuremath{\Xi}\xspace}                 
 \def\PLambda     {\ensuremath{\Lambda}\xspace}                 
 \def\PSigma      {\ensuremath{\Sigma}\xspace}                 
 \def\POmega      {\ensuremath{\Omega}\xspace}                 
 \def\PUpsilon    {\ensuremath{\Upsilon}\xspace}

 \def\PB      {\ensuremath{\mathrm{B}}\xspace}                 
                  
 \def\PD      {\ensuremath{\mathrm{D}}\xspace}

 \def\PJ      {\ensuremath{\mathrm{J}}\xspace}                 
 \def\PK      {\ensuremath{\mathrm{K}}\xspace}

 \def\Pb      {\ensuremath{\mathrm{b}}\xspace}                 
 \def\Pc      {\ensuremath{\mathrm{c}}\xspace}

 \def\Pi      {\ensuremath{\mathrm{i}}\xspace}

 \def\Pp      {\ensuremath{\mathrm{p}}\xspace}

 \def\Ps      {\ensuremath{\mathrm{s}}\xspace}

 \def\thebaroffset{0.0em}
}
{

 \def\Pmu         {\ensuremath{\mu}\xspace}                 
 \def\Pnu         {\ensuremath{\nu}\xspace}                 
                  
 \def\Ppi         {\ensuremath{\pi}\xspace}

 \def\Ptau        {\ensuremath{\tau}\xspace}

 \def\Ppsi        {\ensuremath{\psi}\xspace}                 
                  
 \mathchardef\PDelta="7101
 \mathchardef\PXi="7104
 \mathchardef\PLambda="7103
 \mathchardef\PSigma="7106
 \mathchardef\POmega="710A
 \mathchardef\PUpsilon="7107
                  
 \def\PB      {\ensuremath{B}\xspace}                 
                  
 \def\PD      {\ensuremath{D}\xspace}

 \def\PJ      {\ensuremath{J}\xspace}                 
 \def\PK      {\ensuremath{K}\xspace}

 \def\Pb      {\ensuremath{b}\xspace}                 
 \def\Pc      {\ensuremath{c}\xspace}

 \def\Pi      {\ensuremath{i}\xspace}

 \def\Pp      {\ensuremath{p}\xspace}

 \def\Ps      {\ensuremath{s}\xspace}

 \def\thebaroffset{0.18em}
}
\newcommand{\offsetoverline}[2][\thebaroffset]{\kern #1\overline{\kern -#1 #2}}%

\makeatletter
\ifcase \@ptsize \relax
  \newcommand{\miniscule}{\@setfontsize\miniscule{4}{5}}
\or
  \newcommand{\miniscule}{\@setfontsize\miniscule{5}{6}}
\or
  \newcommand{\miniscule}{\@setfontsize\miniscule{5}{6}}
\fi
\makeatother

\DeclareRobustCommand{\optbar}[1]{\shortstack{{\miniscule (\rule[.5ex]{1.25em}{.18mm})}
  \\ [-.7ex] $#1$}}

\def\mup        {{\ensuremath{\Pmu^+}}\xspace}
\def\mun        {{\ensuremath{\Pmu^-}}\xspace} 

\def\mumu       {{\ensuremath{\Pmu^+\Pmu^-}}\xspace}

\def\taup       {{\ensuremath{\Ptau^+}}\xspace}
\def\taum       {{\ensuremath{\Ptau^-}}\xspace}

\def\neu        {{\ensuremath{\Pnu}}\xspace}
\def\neub       {{\ensuremath{\overline{\Pnu}}}\xspace}

\def\neumb      {{\ensuremath{\neub_\mu}}\xspace}
\def\neut       {{\ensuremath{\neu_\tau}}\xspace}

\def\squark    {{\ensuremath{\Ps}}\xspace}

\def\cquark    {{\ensuremath{\Pc}}\xspace}
\def\cquarkbar {{\ensuremath{\overline \cquark}}\xspace}

\def\bquark    {{\ensuremath{\Pb}}\xspace}

\def\pion   {{\ensuremath{\Ppi}}\xspace}

\def\pip    {{\ensuremath{\pion^+}}\xspace}
\def\pim    {{\ensuremath{\pion^-}}\xspace}

\def\kaon    {{\ensuremath{\PK}}\xspace}

\def\KorKbar {\kern \thebaroffset\optbar{\kern -\thebaroffset \PK}{}\xspace}

\def\Kp      {{\ensuremath{\kaon^+}}\xspace}
\def\Km      {{\ensuremath{\kaon^-}}\xspace}
\def\Kpm     {{\ensuremath{\kaon^\pm}}\xspace}

\def\D       {{\ensuremath{\PD}}\xspace}

\def\DorDbar {\kern \thebaroffset\optbar{\kern -\thebaroffset \PD}\xspace}
\def\Dz      {{\ensuremath{\D^0}}\xspace}

\def\Dp      {{\ensuremath{\D^+}}\xspace}

\def\B       {{\ensuremath{\PB}}\xspace}
\def\Bbar    {{\ensuremath{\offsetoverline{\PB}}}\xspace}

\def\BorBbar {\kern \thebaroffset\optbar{\kern -\thebaroffset \PB}\xspace}
\def\Bz      {{\ensuremath{\B^0}}\xspace}
\def\Bzb     {{\ensuremath{\Bbar{}^0}}\xspace}
\def\Bd      {{\ensuremath{\B^0}}\xspace}
\def\Bdb     {{\ensuremath{\Bbar{}^0}}\xspace}
\def\BdorBdbar {\kern \thebaroffset\optbar{\kern -\thebaroffset \Bd}\xspace}
\def\Bu      {{\ensuremath{\B^+}}\xspace}
\def\Bub     {{\ensuremath{\B^-}}\xspace}
\def\Bp      {{\ensuremath{\Bu}}\xspace}
\def\Bm      {{\ensuremath{\Bub}}\xspace}
\def\Bpm     {{\ensuremath{\B^\pm}}\xspace}

\def\Bs      {{\ensuremath{\B^0_\squark}}\xspace}
\def\Bsb     {{\ensuremath{\Bbar{}^0_\squark}}\xspace}
\def\BsorBsbar {\kern \thebaroffset\optbar{\kern -\thebaroffset \Bs}\xspace}
\def\Bc      {{\ensuremath{\B_\cquark^+}}\xspace}
\def\Bcp     {{\ensuremath{\B_\cquark^+}}\xspace}
\def\Bcm     {{\ensuremath{\B_\cquark^-}}\xspace}

\def\jpsi     {{\ensuremath{{\PJ\mskip -3mu/\mskip -2mu\Ppsi\mskip 2mu}}}\xspace}
\def\psitwos  {{\ensuremath{\Ppsi{(2S)}}}\xspace}

\def\Y#1S{\ensuremath{\PUpsilon{(#1S)}}\xspace}

\def\proton      {{\ensuremath{\Pp}}\xspace}

\def\Lz          {{\ensuremath{\PLambda}}\xspace}

\def\LorLbar     {\kern \thebaroffset\optbar{\kern -\thebaroffset \PLambda}\xspace}

\def\Lc          {{\ensuremath{\Lz^+_\cquark}}\xspace}

\def\Lb           {{\ensuremath{\Lz^0_\bquark}}\xspace}

\def\BF         {{\ensuremath{\mathcal{B}}}\xspace}
\def\BR         {\BF}

\newcommand{\decay}[2]{\ensuremath{#1\!\to #2}\xspace}         

\def\to                 {\ensuremath{\rightarrow}\xspace}

\def\CP                {{\ensuremath{C\!P}}\xspace}

\def\AT#1     {\ensuremath{A_{\mathrm{T}}^{#1}}\xspace}           

\def\C#1      {\ensuremath{\mathcal{C}_{#1}}\xspace}                       
\def\Cp#1     {\ensuremath{\mathcal{C}_{#1}^{'}}\xspace}                    
\def\Ceff#1   {\ensuremath{\mathcal{C}_{#1}^{\mathrm{(eff)}}}\xspace}        
\def\Cpeff#1  {\ensuremath{\mathcal{C}_{#1}^{'\mathrm{(eff)}}}\xspace}       
\def\Ope#1    {\ensuremath{\mathcal{O}_{#1}}\xspace}                       
\def\Opep#1   {\ensuremath{\mathcal{O}_{#1}^{'}}\xspace}                    

\newcommand{\aunit}[1]{\ensuremath{\text{\,#1}}}       

\newcommand{\tev}{\aunit{Te\kern -0.1em V}\xspace}
\newcommand{\gev}{\aunit{Ge\kern -0.1em V}\xspace}
\newcommand{\mev}{\aunit{Me\kern -0.1em V}\xspace}
\newcommand{\kev}{\aunit{ke\kern -0.1em V}\xspace}
\newcommand{\ev}{\aunit{e\kern -0.1em V}\xspace}
\newcommand{\mevc}{\ensuremath{\aunit{Me\kern -0.1em V\!/}c}\xspace}
\newcommand{\gevc}{\ensuremath{\aunit{Ge\kern -0.1em V\!/}c}\xspace}
\newcommand{\mevcc}{\ensuremath{\aunit{Me\kern -0.1em V\!/}c^2}\xspace}
\newcommand{\gevcc}{\ensuremath{\aunit{Ge\kern -0.1em V\!/}c^2}\xspace}

\def\fb   {\ensuremath{\aunit{fb}}\xspace}
\def\invfb   {\ensuremath{\fb^{-1}}\xspace}

\newcommand{\chisq}{\ensuremath{\chi^2}\xspace}

\newcommand{\chisqip}{\ensuremath{\chi^2_{\text{IP}}}\xspace}

\def\gsim{{~\raise.15em\hbox{$>$}\kern-.85em
          \lower.35em\hbox{$\sim$}~}\xspace}
\def\lsim{{~\raise.15em\hbox{$<$}\kern-.85em
          \lower.35em\hbox{$\sim$}~}\xspace}

\def\sqs   {\ensuremath{\protect\sqrt{s}}\xspace}

\def\pt         {\ensuremath{p_{\mathrm{T}}}\xspace}

\def\evtgen     {\mbox{\textsc{EvtGen}}\xspace}

\def\geant      {\mbox{\textsc{Geant4}}\xspace}

\def\photos     {\mbox{\textsc{Photos}}\xspace}

\def\pythia     {\mbox{\textsc{Pythia}}\xspace}

\xspace

\def\tell1  {TELL1\xspace}
\def\ukl1   {UKL1\xspace}

\newcommand{\etal}{\mbox{\itshape et al.}\xspace}

%% file: title-LHCb-PAPER.tex

\begin{titlepage}
\pagenumbering{roman}

\vspace*{-1.5cm}
\centerline{\large EUROPEAN ORGANIZATION FOR NUCLEAR RESEARCH (CERN)}
\vspace*{1.5cm}
\noindent
\begin{tabular*}{\linewidth}{lc@{\extracolsep{\fill}}r@{\extracolsep{0pt}}}
\ifthenelse{\boolean{pdflatex}}
{\vspace*{-1.5cm}\mbox{\!\!\!\includegraphics[width=.14\textwidth]{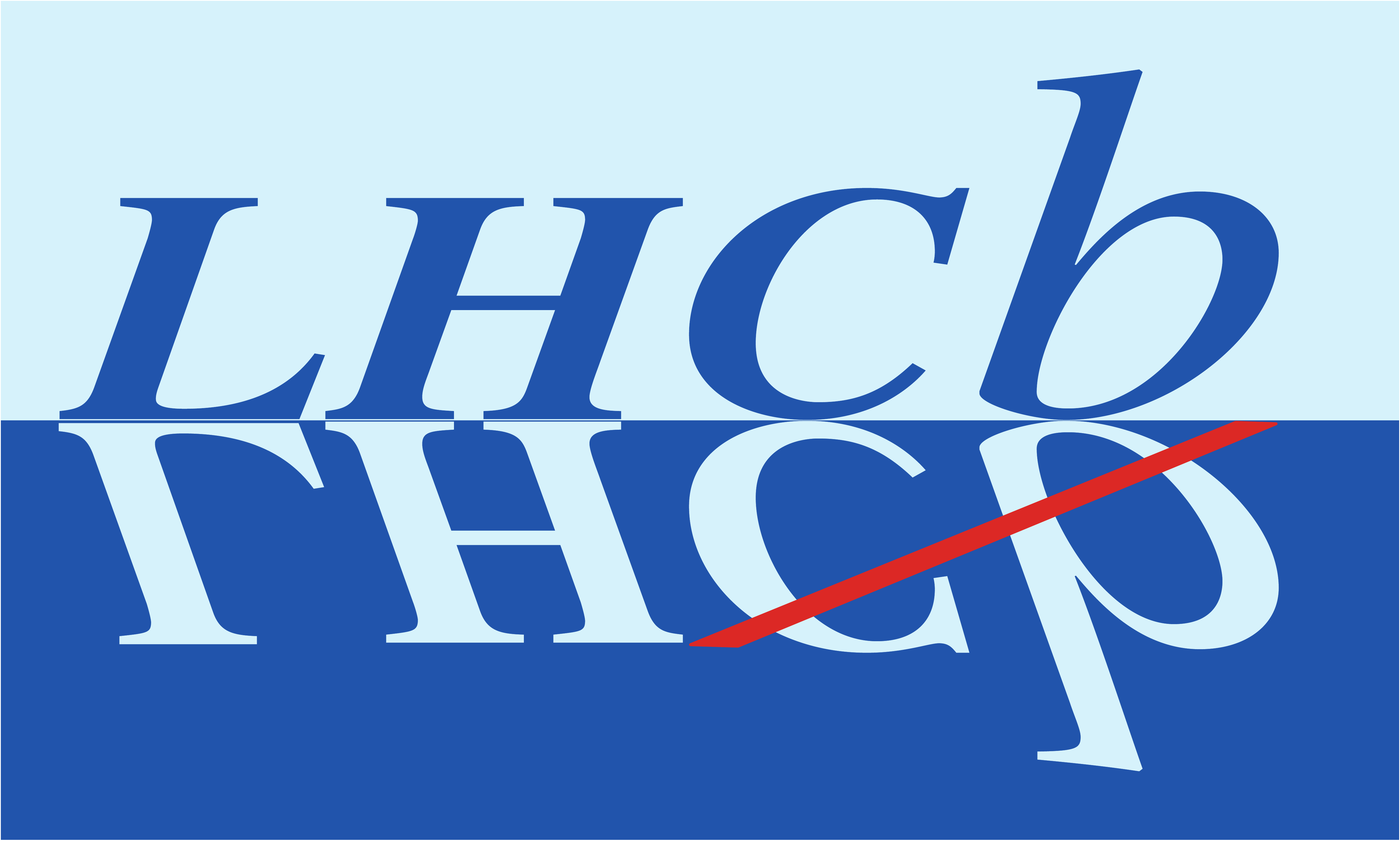}} & &}%
{\vspace*{-1.2cm}\mbox{\!\!\!\includegraphics[width=.12\textwidth]{lhcb-logo.eps}} & &}%
\\
 & & CERN-EP-2019-216 \\  
 & & LHCb-PAPER-2019-033\\  
 & & \today \\ 
 & & \\
\end{tabular*}

\vspace*{4.0cm}

{\normalfont\bfseries\boldmath\huge
\begin{center}
  \papertitle 
\end{center}
}

\vspace*{2.0cm}

\begin{center}
\paperauthors\footnote{Authors are listed at the end of this paper.}
\end{center}

\vspace{\fill}

\begin{abstract}
  \noindent

   The production fraction of  the \Bcm meson with respect to the sum of $\Bub$ and $\Bdb$ mesons is measured in both 7 and 13\tev center-of-mass energy $pp$ collisions produced by the Large Hadron Collider (LHC), using the LHCb detector. The rate, approximately 3.7 per mille,  does not change with energy, but shows a transverse momentum dependence. The $\Bcm-\Bcp$ production asymmetry is also measured, and is consistent with zero within the determined statistical and systematic uncertainties of a few percent.
    
\end{abstract}

\vspace*{2.0cm}

\begin{center}
  Published in Phys. Rev. D100 (2019) 112006 
  \end{center}

\vspace{\fill}

{\footnotesize 
\centerline{\copyright~\papercopyright. \href{\paperlicenceurl}{\paperlicence}.}}
\vspace*{2mm}

\end{titlepage}


\newpage
\setcounter{page}{2}
\mbox{~}
%
%
%
%

\cleardoublepage

%% file: 1introduction.tex
\section{Introduction}
\label{sec:introduction}

The \Bcm meson is a bound state containing  a $b$ quark with a $\overline{c}$ quark.\footnote{The mention of a particular state implies the use of the charge-conjugate state as well, except when discussing production asymmetries.} It has the largest mass of any two differently flavored quarks in a mesonic ground state.  Studies of its production or determination of individual decay widths require measurements of its branching fractions to exclusive final states. 
 Since the branching fractions of some decay modes of \Bm and \Bzb mesons are accurately known, we determine the 
ratio of \Bcm meson production relative to the sum of \Bm and \Bzb mesons. Here we use techniques similar to those employed for the measurement of \Bsb meson and \Lb baryon fractions \cite{LHCb-PAPER-2018-050}. In that paper use is made of the fact that the semileptonic widths of all $b$-flavored hadrons with light and strange quarks are equal. However, both the $b$ and $c$ quarks can decay, rendering that concept inapplicable. Instead we rely on theoretical predictions of the semileptonic decay branching fraction ${\cal{B}}(\Bcmu)$. Currently, only the relative production cross-section times the branching
fraction of either the $\Bcmu$ or $\Bcm\to\jpsi\pim$ modes have been measured by the \cdf
\cite{Abe:1998wi,Abe:1998fb}, \lhcb
\cite{LHCb-PAPER-2012-028,LHCb-PAPER-2014-050} and \cms \cite{CMS:2013pua,CMS:2018ynq} experiments.

The $\Bcm$ meson production fraction ($f_c$) relative to the sum of $\Bzb$ ($f_d$) and $\Bm$ ($f_u$) mesons is defined as
\begin{equation}\label{eq:frac}
R_c=\frac{f_c}{f_u+f_d}\equiv{n_{\rm cor}(\Bcmu)\over n_{\rm cor}(B\to\Dz X\mu^-\overline{\nu})+n_{\rm cor}(B\to\Dp X\mu^-\overline{\nu})}\cdot{\left<\BR_{\rm sl}\right>\over\BR(\Bcmu)},
\end{equation}
where $n_{\rm cor}$ refers to the efficiency and branching fraction corrected number of signal events. The modes containing $D^0$ and $D^+$ mesons are also corrected for cross-feeds with \Bsb and \Lb decays. The determination of the corrected yields of the $B\to DX\mu^-\overline{\nu}$ decays follows our previous measurement strategy in Ref.~\cite{LHCb-PAPER-2018-050} where the equations relating the fractions to the corrected yields, including cross-feed contributions, are given. We also correct for the 0.4\% effect of doubly-Cabibbo-suppressed decays and $D^0$ mixing. The relevant hadron branching fractions are listed in Table \ref{tab:charmdecay}.
The average semileptonic branching fractions of \Bzb and \Bm, 
$\left<\cal{B}_{\rm sl}\right>=(10.70\pm0.19)\%$  is found by averaging  measurements from the \cleo \cite{Mahmood:2004kq}, \babar \cite{Aubert:2006au} and \belle \cite{Urquijo:2006wd} experiments, detailed in Ref.~\cite{LHCb-PAPER-2016-031}. Since only $b\to c\mun\neumb$ modes are detected in this analysis, a correction for the small $b\to u\mun\neumb$ rate of 1\% is applied to the denominator of Eq.~\ref{eq:frac}. 
 \begin{table}[htp]
\centering
\caption{Charm and charmonium branching fractions for the decay modes used in this analysis. }
\vspace{0.2cm}
\begin{tabular}{lcl}
\hline\hline
Particle and decay & $\BR(\%)$  & Source \\\hline
$\Dz\to\Km\pip$ & $3.93\pm0.03$ & PDG average\cite{PDG2018}\\
$\Dp\to\Km\pip\pip$ & $9.22\pm0.17$  & CLEO III\cite{Bonvicini:2013vxi}\\
$\jpsi\to\mumu$ & $5.96\pm0.03$ & PDG average\cite{PDG2018}\\
\hline\hline 
\end{tabular}
\label{tab:charmdecay}
\end{table}

The dominant production mechanism for \Bcm mesons is gluon-gluon fusion, $gg\to\Bcm+\bar{b}+c$. Non-relativistic quantum chromodynamics is used along with fragmentation functions to predict cross-sections as functions of transverse momentum (\pt) and pseudorapidity ($\eta$). The literature is nicely summarized in Ref.~\cite{Brambilla:2010cs}. We define $H_b$ to refer to $B_c$, $\Bzb$, and $B^-$ mesons, while $H_c$ refers to $D^0$ and $D^+$ mesons.

 In this analysis $\eta$ is determined by measuring the angle of the $B$ meson with respect to the beam direction by using the positions of the primary $pp$ interaction vertex (PV) and the $B$ meson decay point into either $\jpsi\mu^-$, $D^0\mu^-$, or $D^+\mu^-$. The transverse momentum initially refers to the vector sum of the charmed-hadron and $\mu^-$ momentum transverse to the proton beams. However, the results are re-interpreted in terms of the $H_b$ meson $\pt(H_b)$ by simulating and correcting the effects of the missing momenta.

The production  asymmetry between $\Bcm$ and $\Bcp$ mesons, which should be small, is defined as
\begin{eqnarray}
   \aprod\equiv{\sigma(\Bcm)-\sigma(\Bcp)\over\sigma(\Bcm)+\sigma(\Bcp)}=\araw-a_{\rm
   det},
\end{eqnarray}
where $\araw$ and $a_{\rm det}$ are the asymmetries in the signal yields and
the efficiencies of $\Bcm$ and $\Bcp$ detection, respectively. The \CP
asymmetry in the $\Bcm\to\jpsi\mu^-\overline{\nu}$ decay is assumed to be zero in this analysis.

%% file: 2bfprediction.tex
\label{sec:analysis}

\begin{table}[b]
\centering
\caption{Branching fractions predictions (\%). The $\Bcm$ lifetime is taken as 0.507 \!ps \cite{PDG2018}. The value for the semileptonic decays of the \Bcm meson, ${\cal{B}}^c_{\rm sl}$, is derived by summing the $\jpsi\mu^-\overline{\nu}$ and $\eta_c\mu^-\overline{\nu}$ individual predictions with the average predictions of 0.1\% for $\psi(2S)\mu^-\overline{\nu}$, the sum of $\chi_{c0,1,2}\mu^-\overline{\nu}$ as 0.6\%, and 0.3\% for $h_c\mu^-\nu$. In the one case where $\eta_c\mu^-\overline{\nu}$ was not predicted averages from other measurements are used.}

\vspace{0.2cm}
\begin{tabular}{cccccc|c}
\hline
Ref.$\backslash$Mode&$\jpsi\mu^-\overline{\nu}$&$\eta_c\mu^-\overline{\nu}$&$\psi(2S)\mu^-\overline{\nu}$ &$ \chi_{c0,1,2}\mu^-\overline{\nu}$&$h_c\mu^-\nu$&${\cal{B}}^c_{\rm sl}$\\\hline
 \cite{Rui:2016opu} &6.4&5.0&1.3&   & & 13.6\\
 \cite{Ebert:2010zu}&     &      &     & 0.5&&\\
 \cite{Ebert:2003cn}&1.4&0.5&    &  &&\phantom{0}2.9    \\
 \cite{Qiao:2012vt}  &7.5&2.4&    &    && 10.9 \\
 \cite{Chang:2014jca}&1.9&0.6&0.1& &&\phantom{0}3.5   \\
 \cite{Ivanov:2006ni}&2.3&0.9&  & 0.8& &\phantom{0}4.2\\
 \cite{Huang:2007kb}&2.7&1.8&  &    & &\phantom{0}5.5\\
 \cite{Wang:2008xt}&1.6&0.8&   &     &&\phantom{0}3.4\\
 \cite{Hernandez:2006gt}&1.7&0.5&  &0.6& &\phantom{0}3.3\\
 \cite{Colangelo:1999zn}&1.7&0.2&   &    & &\phantom{0}2.9\\
 \cite{Gouz:2002kk}&1.9& 0.8  &0.1&  &&\phantom{0}3.7\\
 \cite{AbdElHady:1999xh}&2.3&0.9&   &   &&\phantom{0}4.2\\
 \cite{Kiselev:2002vz}&2.2&0.8&0.1&    && \phantom{0}4.0\\
 \cite{Chang:1992pt}&2.6&     &0.1&1.1& &\phantom{0}4.2\\
 \cite{Ivanov:2000aj}&2.5&1.1&    & &&\phantom{0}4.6\\
 \cite{Scora:1995ty}&1.3&0.8&0.2&  &&\phantom{0}3.1\\
 \cite{Anisimov:1998xv}&1.4&0.7&   &  &&\phantom{0}3.1\\
\cite{Wang:2018duy}&1.5&0.7& & 0.5&0.3&\phantom{0}3.2 \\
\cite{Geng:2018qrl}& 1.9& 0.6&0.1&0.3&0.3&\phantom{0}3.5\\
\cite{Leljak:2019eyw}& 2.2& 0.8& & & & \phantom{0}4.0\\
\hline\hline 
\end{tabular}
\label{tab:models}
\end{table}

 The branching fraction predictions from various models of semileptonic \Bcm decays are listed in
Table~\ref{tab:models}. For ${\cal{B}}(\Bcm\to
\jpsi\mu^-\overline{\nu})$ they range from 1.4 to 7.5\%, which is quite a large interval.
Branching fractions for other modes are also listed where available. We use the Z expansion fit results from Ref.~\cite{Wang:2018duy}, and the method II results for Ref.~\cite{Geng:2018qrl}.

Some restrictions
on models are possible by comparing to lighter $B$ meson decays. Since the
inclusive semileptonic branching fraction for these decays, ${\cal{B}_{\rm sl}}$, is
about 10.5\% and the $\Bcm$ lifetime, $\tau_{B_c}$, is 1/3 that of the $\Bzb$,\footnote{This is evident since ${\cal{B}}^c_{\rm sl}= \Gamma_{\rm sl}\cdot \tau_{B_c}$, and $\Gamma_{\rm sl}$ is approximately the same for all $b$-hadron species. We use natural units where $c=\hbar=1.$}  we disregard models that predict 10\% or larger values for ${\cal{B}}^c_{\rm sl}$ of the \Bcm.
This excludes from consideration the models of
Refs.~\cite{Rui:2016opu} and \cite{Qiao:2012vt}.  
The average model prediction is then ${\cal{B}}(\Bcm\to \jpsi\mu^-\overline{\nu})=1.95$\%. The standard deviation is 0.46\%, which we use to estimate the systematic uncertainty on the model variation. Results of our measurement without using this branching fraction are also quoted.

%% file: detector.tex
\section{Detector, trigger and simulation}
\label{sec:Detector}
The LHCb 
detector~\cite{LHCb-DP-2008-001,LHCb-DP-2014-002} is a
single-arm forward spectrometer covering the pseudorapidity range $2 < \eta < 5$, designed for
the study of particles containing \bquark\ or \cquark\ quarks. The detector elements that are particularly
relevant to this analysis are: a silicon-strip vertex detector surrounding the $pp$ interaction
region that allows \cquark\ and \bquark\ hadrons to be identified from their characteristically long
flight distance; a tracking system that provides a measurement of the momentum, $p$, of charged
particles; two ring-imaging Cherenkov detectors that are able to discriminate between
different species of charged hadrons; and a downstream system of iron interspersed with chambers is used to identify muons.

 The magnetic field deflects positively and negatively charged particles in opposite
directions and this can lead to detection asymmetries. Periodically
reversing the magnetic field polarity throughout the data taking almost cancels
the effect. The configuration with the magnetic field pointing upwards (downwards) bends positively (negatively) charged particles
in the horizontal plane towards the centre of the LHC ring.
This analysis uses data collected in 2011 (7\tev) and 2016 (13\tev) where appropriate triggers are available.  The data taking was split between magnetic field up and down configurations.
In the 2011 data 0.6\invfb (0.4\invfb) were collected with the field pointing up (down), while in 2016 the split was 0.9\invfb with field up and 0.8\invfb with field down. 

The trigger~\cite{LHCb-DP-2012-004} consists of a
hardware stage, based on information from the calorimeter and muon
systems, followed by a software stage, in which all charged particles
with $\pt>500\,(300)\mev$ are reconstructed for 2011\,(2016) data.

Separate hardware triggers are used for the $\jpsi\mu^-$ and $H_c$ samples.
For the former we require a $\mu^+\mu^-$ pair. For the latter, we require a single muon with large \pt  for the 7\tev data as used in Ref.~\cite{LHCb-PAPER-2011-018}. For the 13\tev data, the single muon trigger was not available, therefore at the hardware trigger stage, events are required to have a hadron, photon or electron transverse energy greater than approximately 3.5 GeV in the calorimeters.
The software trigger requires a two-, three- or four-track
  secondary vertex with a significant displacement from any primary
  $pp$ interaction vertex as described in Ref.~\cite{LHCb-PAPER-2018-050}. 
  At least one charged particle
  must have $\pt > 1.6\gev$ and be
  inconsistent with originating from a PV.
  A multivariate algorithm~\cite{BBDT} is used for
  the identification of secondary vertices consistent with the decay
  of a \bquark hadron.
  
   Simulation is required to model the effects of the detector acceptance and the
  imposed selection requirements.
  In the simulation, $pp$ collisions are generated using
\pythia~\cite{Sjostrand:2006za,*Sjostrand:2007gs} 
 with a specific \lhcb
configuration~\cite{LHCb-PROC-2010-056}.  Decays of unstable particles
are described by \evtgen~\cite{Lange:2001uf}, in which final-state
radiation is generated using \photos~\cite{Golonka:2005pn}. The
interaction of the generated particles with the detector, and its response,
are implemented using the \geant
toolkit~\cite{Allison:2006ve, *Agostinelli:2002hh} as described in
Ref.~\cite{LHCb-PROC-2011-006}.

%% file: acknowledgements.tex
\section*{Acknowledgements}
%
%
\noindent We express our gratitude to our colleagues in the CERN
accelerator departments for the excellent performance of the LHC. We
thank the technical and administrative staff at the LHCb
institutes.
We acknowledge support from CERN and from the national agencies:
CAPES, CNPq, FAPERJ and FINEP (Brazil); 
MOST and NSFC (China); 
CNRS/IN2P3 (France); 
BMBF, DFG and MPG (Germany); 
INFN (Italy); 
NWO (Netherlands); 
MNiSW and NCN (Poland); 
MEN/IFA (Romania); 
MSHE (Russia); 
MinECo (Spain); 
SNSF and SER (Switzerland); 
NASU (Ukraine); 
STFC (United Kingdom); 
DOE NP and NSF (USA).
We acknowledge the computing resources that are provided by CERN, IN2P3
(France), KIT and DESY (Germany), INFN (Italy), SURF (Netherlands),
PIC (Spain), GridPP (United Kingdom), RRCKI and Yandex
LLC (Russia), CSCS (Switzerland), IFIN-HH (Romania), CBPF (Brazil),
PL-GRID (Poland) and OSC (USA).
We are indebted to the communities behind the multiple open-source
software packages on which we depend.
Individual groups or members have received support from
AvH Foundation (Germany);
EPLANET, Marie Sk\l{}odowska-Curie Actions and ERC (European Union);
ANR, Labex P2IO and OCEVU, and R\'{e}gion Auvergne-Rh\^{o}ne-Alpes (France);
Key Research Program of Frontier Sciences of CAS, CAS PIFI, and the Thousand Talents Program (China);
RFBR, RSF and Yandex LLC (Russia);
GVA, XuntaGal and GENCAT (Spain);
the Royal Society
and the Leverhulme Trust (United Kingdom).

%% file: Bcfrac-prl-v5.bbl
\ifx\mcitethebibliography\mciteundefinedmacro
\PackageError{LHCb.bst}{mciteplus.sty has not been loaded}
{This bibstyle requires the use of the mciteplus package.}\fi
\providecommand{\href}[2]{#2}

%% file: LHCb_Authorship_10-Sep-2019.tex
\centerline
{\large\bf LHCb collaboration}
\begin
{flushleft}
\small
R.~Aaij$^{31}$,
C.~Abell{\'a}n~Beteta$^{49}$,
T.~Ackernley$^{59}$,
B.~Adeva$^{45}$,
M.~Adinolfi$^{53}$,
H.~Afsharnia$^{9}$,
C.A.~Aidala$^{79}$,
S.~Aiola$^{25}$,
Z.~Ajaltouni$^{9}$,
S.~Akar$^{64}$
% ...

{\footnotesize \it

% ...
$ ^{9}$Universit{\'e} Clermont Auvergne, CNRS/IN2P3, LPC, Clermont-Ferrand, France\\
% ...
$ ^{25}$INFN Sezione di Milano, Milano, Italy\\
% ...
$ ^{31}$Nikhef National Institute for Subatomic Physics, Amsterdam, Netherlands\\
% ...
$ ^{45}$Instituto Galego de F{\'\i}sica de Altas Enerx{\'\i}as (IGFAE), Universidade de Santiago de Compostela, Santiago de Compostela, Spain\\
% ...
$ ^{49}$Physik-Institut, Universit{\"a}t Z{\"u}rich, Z{\"u}rich, Switzerland\\
% ...
$ ^{53}$H.H. Wills Physics Laboratory, University of Bristol, Bristol, United Kingdom\\
% ...
$ ^{59}$Oliver Lodge Laboratory, University of Liverpool, Liverpool, United Kingdom\\
% ...
$ ^{64}$University of Cincinnati, Cincinnati, OH, United States\\
% ...
$ ^{67}$Syracuse University, Syracuse, NY, United States\\
% ...
$ ^{79}$University of Michigan, Ann Arbor, United States, associated to $^{67}$\\
\bigskip
% ...
}
\end{flushleft}